\documentclass{ar-1col-S2O}
\usepackage{aas_macros}
\usepackage{amsmath}
\usepackage{amssymb}

\usepackage[comma]{natbib}
\usepackage{url}
\usepackage{verbatim}
\usepackage{graphicx}
\graphicspath{{figs/}}

\jyear{2025}

\def\chandra{{\it Chandra}}
\def\xmm{{\it XMM-Newton}}

\def\swift{{\it Swift}}
\def\nustar{{\it NuSTAR}}
\def\nicer{{\it NICER}}
\def\xrism{{\it XRISM}}
\def\ixpe{{\it IXPE}}
\def\xillver{{\sc xillver}}
\def\relxill{{\sc relxill}}

\def\spin{$a_\star$}
\def\msun{$M_{\odot}$}
\def\mdot{$\dot M$}

\def\risco{$R_{\rm ISCO}$}

\begin{document}

\markboth{Kara \& Garc\'ia}{X-rays from Accreting Black Holes}

\title{Supermassive Black Holes in X-rays: From Standard Accretion to Extreme Transients}

\author{Erin Kara$^1$ and Javier Garc\'ia$^2$
\affil{$^1$MIT; email: ekara@mit.edu}
\affil{$^2$NASA Goddard Space Flight Center; email: javier.a.garciamartinez@nasa.gov}}

\begin{abstract}

X-rays are a critical wavelength for understanding supermassive black holes (SMBHs). X-rays probe the inner accretion flow, closest to the event horizon, where gas inspirals, releasing energy and driving black hole growth. 
This region also governs the launching of outflows and jets that regulate galaxy evolution and link SMBH growth to their host galaxies.\\

This review focuses on X-ray observations of SMBHs, through ``standard accretion'' in persistent Active Galactic Nuclei (AGN) and in extreme transient events, such as Tidal Disruption Events (TDEs), Changing-Look AGN (CLAGN), and Quasi-Periodic Eruptions (QPEs). We describe the X-ray spectral and variability properties of AGN, and the observational techniques that probe the inner accretion flow. Through understanding the phenomenology and accretion physics in standard, individual AGN, we can better probe more exotic phenomenon, including binary SMBH mergers, or Extreme Mass Ratio Inspirals (EMRIs).\\

In this review, the reader will discover:

\parbox{10cm}{
\begin{itemize}
    \item X-ray variability on timescales from minutes to hours traces accretion near the event horizon.
    \item{X-rays can measure the black hole mass, spin and accretion flow geometry and dynamics.}
    \item{In transients like TDEs, X-rays probe the newly formed accretion disk that feeds the black hole.}
    \item{QPEs are posited to be EMRIs orbiting accreting SMBHs that would emit low-frequency gravitational waves.}
    \item Future X-ray, time-domain and multi-messenger surveys will revolutionize our understanding of SMBH growth.
\end{itemize}
}

\end{abstract}

\maketitle

\tableofcontents

\section{INTRODUCTION}

\subsection{A Brief History of Supermassive Black Holes in X-rays}
\label{sec:intro}

When the astronomical object 3C~273 was first detected \citep{Edge1959}, to most optical telescopes it looked like a star in our Galaxy (and was so-termed a ``Quasi Stellar Object" or QSO). But in 1963, astronomers realized that the spectra were highly redshifted, implying a distance of 750 Mpc ($z=0.158$), and thus the bright 3C~273 was, in fact, the most luminous object ever seen \citep{Hazard1963, Schmidt1963}. In addition, the optical and radio emission from 3C~273 was variable--on timescales of years, to as short as days \citep{Smith1963, Dent1965}, indicating that the size of the emitting region was a few light-days across. In other words, these QSOs were producing more radiation than a trillion stars from a region only the size of our Solar System! Not long-after, theorists inferred that accretion onto supermassive black holes could produce such high luminosities from such compact regions \citep{Salpeter1964}. Gas within the gravitational sphere of influence of the black hole will collide, and thanks to the conservation of angular momentum, will eventually form a rotating accretion disk that emits blackbody radiation at optical/UV wavelengths \citep{Lynden-Bell1969, Bardeen1970}. {\em Through the combination of spectral and timing observations, the field of supermassive black hole accretion and growth was born.} 

Concurrent with the discovery of the first QSO (now used interchangeably with the term Quasars), the first X-ray sounding rockets were taking flight \citep{Giacconi1962}. It did not take long to discover that the quasar 3C~273 was also bright in X-rays \citep{Bowyer1970}, and beyond that, X-rays appeared commonplace in all AGN, even in lower mass, nearby Seyferts \citep{Elvis1978}. Indeed, today, it is widely recognized that {\em X-rays are the most efficient means for unambiguously identifying AGN} because: (1) X-rays are ubiquitous in AGN and, (2) X-rays are not subject to ``contamination" from starlight, as is often the case at longer wavelengths \citep{Mushotzky1993}.

Repeated measurements led quickly to the discovery that AGN vary much more rapidly in X-rays than they do in optical \citep{Winkler1975, Marshall1981}, sometimes on timescales as short as a few minutes \citep{McHardy1981}. This showed that the X-ray emission region arises deep within the nucleus, probing the very central engine of the AGN. While the $\sim10^{4-6}$\,K accretion disk could be hot enough to explain the optical emission, the X-rays must come from an even hotter ($\sim10^9$\,K), and more compact plasma, that we now refer to as the ``X-ray corona". This definition is based on the spectral and timing properties observed, while the details of its geometry, location, dynamics, and kinematics are still a matter of research.
\begin{marginnote}[]
    \entry{X-ray Corona}{
    Billion degree optically-thin plasma where the X-ray continuum originates.
    }
\end{marginnote}

Beyond the broadband continuum (well-described in the 2--10 keV band as a powerlaw with photon index $\Gamma\sim 1.5-2$), \citet{Mushotzky1978} discovered distinctive features at $\sim6.4$~keV due to the fluorescence of iron as electron transitions from the inner-most atomic shell (K-shell in spectroscopic notation). This indicates that the central, hot corona irradiates colder gas that flows around the AGN. This Fe K line is a hallmark of the X-ray spectrum and is nearly ubiquitous in local Seyferts \citep{Pounds1989}. \citet{Fabian1989} posited that the fluorescence-emitting gas could be the accretion disk itself, and thus by measuring Doppler and gravitational shifts in the Fe K line, one can estimate properties of the disk, including its inclination, and how far the disk extends towards the black hole (an indicator of the spin; see Section~\ref{sec:reflection} for details). {\em And thus by combining X-ray spectral and timing information of AGN, one can measure fundamental properties of the black hole (its mass and spin), and can uniquely probe the regions closest to the event horizon, where most of the accretion power is released}. Paraphrasing Prof. Andy Fabian:

\begin{extract}
    ``It is quite remarkable that the X-ray corona, which is one of the brightest regions in astronomical objects, is coincidentally located right next to the darkest objects in the Universe, black holes."
\end{extract}
\subsection{A note from the Authors}

The last Annual Reviews dedicated specifically to the X-ray spectral and timing properties of AGN was published over thirty years ago \citep{Mushotzky1993}, before the launch of our two X-ray flagship missions, {\chandra} and {\xmm}, and before even the launch of the first X-ray CCDs on the {\it ASCA} mission.
\begin{marginnote}[]
    \entry{CCD}{
    {\it Charge-Coupled Device}; a commonly used detector that converts light into electronic impulses.
    }
\end{marginnote}
The past decade has seen a true revolution in our understanding of the inner accretion flows of AGN, thanks, in large part, to these two flagships and a suite of small, dedicated X-ray telescopes each opening a new and unique parameter space (e.g.; \swift, \nustar, {\it AstroSAT}, \nicer, {\it Insight-HXMT}, \ixpe, and \xrism).

Beyond advancements specific to the X-ray band, the field of black hole accretion is seeing a renaissance in the wider astronomical community in the last 5--10 years, thanks to the advent of wide-field, time-domain surveys across the electromagnetic spectrum. These surveys monitor hundreds of thousands of AGN at unprecedented cadence, revealing the secrets AGN were keeping while we weren't watching. Time domain surveys are changing what we thought we understood about standard AGN activity, and thus, evolving our picture of how supermassive black holes grow and affect their environments. 

In writing this Annual Review, we aim to describe what we know about ``Standard AGN", and how we can use X-ray and multi-wavelength properties to understand their mass, spin, and accretion flow properties (Chapters~\ref{chap:accretion} and \ref{chap:AGN}). {\it It is only if we truly understand ``Standard AGN'' that we will be able to make sense of the new exotic transients being discovered through time-domain and multi-messenger surveys (Chapter~\ref{chap:variability}).}

In a short review, it is impossible to cover all of AGN in X-rays. Admittedly, we limit the present discussion to our areas of expertise, in the X-ray spectral and timing observations of luminous ($L > 0.01~L_{\mathrm{Edd}}$), low-obscuration ($N_{H}<10^{23}$~cm$^{-2}$), largely radio-quiet ($L_{\mathrm{1400~MHz}}< 30~\mu$Jy), local ($z<1$) AGN. In other words, {\em we focus this review on observations of the central engines of supermassive black holes (SMBHs) at the time in their evolution of rapid growth}. For readers interested in these other topics, we refer to other recent Annual Reviews: for simulations, see \cite{Davis+2020}; for AGN jets, see \cite{blandford2019}; for AGN outflows, see \cite{King2015}; for obscured AGN, see \cite{Hickox2018}, for low-luminosity AGN, see \cite{Yuan2014}; for AGN feedback, see \cite{Fabian2012}. 

We aim for this review to not only be a tool for X-ray astronomers, but also for the broader community working on time-domain and multi-messenger astrophysics, MHD
simulations of black holes, and high-$z$ quasars with {\it JWST}, who would like to understand better what to expect from X-ray observations of standard AGN, and how extreme accretion deviates from our standard picture. We hope this review will build connections across several sub-fields of black hole accretion by highlighting what we have collectively learned in recent years, thus facilitating future discoveries.
\begin{marginnote}[]
    \entry{MHD}{
    {\it Magneto-Hydro-Dynamics}; the time-dependent solution of fluid equations in magnetic fields.
    }
\end{marginnote}
%

\section{ACCRETION ONTO COMPACT OBJECTS}\label{chap:accretion}

The high-energy X-rays observed from AGN are not directly emitted by the accretion disk, but the radiation is primarily powered by accretion processes onto the black hole, which we review here (for details on the X-ray emitting region, see Sec.~\ref{sec:corona}). This Section begins with an overview of accretion theory, used to describe flows influenced by the gravitational potential of black holes, neutron stars, and, to some extent, white dwarfs. Readers seeking a more detailed discussion are encouraged to consult works such as \cite{Armitage+2022,Abramowicz+2013}, and \cite{Davis+2020}.

While the fundamentals of accretion theory discussed here are simplified, they form the basis of much of the historical interpretation of AGN observations. Over the past decade, advancements in computing and simulations have enabled more realistic models of black hole accretion flows, that include 3D effects and radiation. These theoretical developments allows us to start making comparison to observations, and present both opportunities and challenges for theorists to refine models that match the requirements of increasingly sophisticated observational data.

\subsection{The basics of accretion theory}

The two most fundamental quantities to describe an accretion disk around
a black hole are the mass $M$ and the rate at which mass from the accretion disk falls into the black hole, $\dot M$.
The total mass determines the size scale of the system through the
{\it Schwarzschild radius}, which coincides with the radius of the event horizon in a
non-spinning black hole:
\begin{equation}\label{eq:rs}
R_s = 2 R_g = \frac{2GM}{c^2} \simeq 3 \left( \frac{M}{M_{\odot}} \right) \mathrm{km}
\end{equation}
where $R_g$ is the {\it gravitational radius}, $G$ is the Newtonian gravitational constant and $c$ is the speed of light. To give a sense of how compact black holes can be, a $10^8$ solar masses black hole has a horizon of $R_s \sim 1$\,AU, fitting comfortably within the Earth's orbit.

The extraction of gravitational potential energy through accretion is likely the most efficient and powerful mechanism for the production of high-energy radiation. This can be shown by simple order-of-magnitude estimates. For a body of mass $M$ and radius $R$, the gravitational potential energy released by accretion of a particle of mass $m$ onto its surface is
\begin{equation}\label{egrav}
\Delta E_{acc} = \frac{GMm}{R},
\end{equation}
In the interior of stars, the burning of hydrogen through nuclear fusion releases no more than 1\% of the rest-mass energy, $\Delta E_{nuc} = 0.01mc^2$. Dividing these two one gets
\begin{equation}
    \frac{\Delta E_{acc}}{\Delta E_{nuc}} = 10^2 \left( \frac{R_g}{R} \right).
\end{equation}

Thus, for a Schwarzschild black hole ($R=2R_g$), the energy released from accetion is $\sim50$ times that released from nuclear burning.
More detailed estimates predict that up to 40\% of the rest mass of the accreted material can be emitted as radiation \citep{mcc06}.

The accretion rate defines the total luminosity
of the system
\begin{equation}\label{eq:eta}
L = \eta \dot M c^2
\end{equation}
Here $\eta$ is the efficiency of converting the rest-mass energy of the infalling material into radiation. Importantly, whatever is not converted into luminosity ($(1-\eta)Mc^2$) is accreted onto the black hole, increasing its mass. 

There exist a critical accretion rate at which the outward radiation pressure stops the inward accretion of material. This situation is known as the {\it Eddington limit}. Assuming spherical accretion of hydrogen gas onto a compact object of mass $M$, if the gravitational attraction at a radius $r$ is balanced with the radiation pressure:
\begin{equation}
\frac{GMm_p}{R^2} = \frac{L \sigma_T}{4\pi R^2 c}
\end{equation}
where $m_p$ is the mass of the proton (much larger than the electron mass), and $\sigma_T = 8\pi e^4/3m_e^2c^4\approx6.64\times 10^{-25}$\,cm$^2$ is the Thomson cross section for electron scattering, then the {\it Eddington Luminosity} is
\begin{equation}
L_{Edd} = \frac{4\pi GMm_pc}{\sigma_T} \simeq 1.26 \times 10^{44} \left(
\frac{M}{10^6 M_{\odot}}\right) \mathrm{erg \ \ s^{-1}}.
\end{equation}
Although the assumption of spherical symmetry is likely an oversimplification, observational evidence suggest that most accreting systems appear to largely obey it. The Eddington luminosity also implies a critical accretion rate given by
\begin{equation}
\dot M_{Edd} = \frac{L_{Edd}}{\eta c^2} \simeq 0.03 \left(\frac{M}
{10^6 M_{\odot}} \right) \left( \frac{0.08}{\eta}\right) M_{\odot} \mathrm{\ \ yr}^{-1}.
\end{equation}
The efficiency of accretion can be estimated by comparing the binding energy of a particle at the {\it innermost stable circular orbit} (ISCO) with its rest mass energy $E=mc^2$.

For a non-spinning black hole \risco$ = 6R_g$ \citep{kap49}, and using the Newtonian expression for the binding energy
\begin{equation}
E_b = \frac{GMm}{2R_g} = \frac{mc^2}{12}
\end{equation}
then
\begin{equation}
\eta = \frac{1}{12} = 0.08
\end{equation}
\begin{marginnote}[]
    \entry{$R_\mathrm{ISCO}$}{
    Radius of the inner-most circular orbit; is the smallest radius at which a test particle can sustain stable circular orbits around a black hole (or neutron star).}
\end{marginnote}
A more detailed calculation taking into account the space-time curvature close to the black hole reduces the binding energy to $\eta = 0.057$ \citep{Salpeter1964}.

\subsection{Standard Accretion Disk Theory}

The most standard semi-analytical treatment of the theory of accretion disks is that by \cite{Shakura1973} and \cite{nov73}, the latter including a fully relativistic framework. Here we simply mention some of the most important derivations, and for simplicity we will follow the classical approach of \cite{Shakura1973} (hereafter, SS73).

The standard theory considers a geometrically thin accretion disk (i.e. $H \ll R$, where $H$ is the scale height at radius $R$), such that the gas particles follow Keplerian orbits at each radius and the inward radial speed is much smaller than the rotation velocity. The internal viscosity of the gas produces a loss of energy and angular momentum of the particles. This model prescribes that the viscosity $\nu$ in the gas is given by
\begin{equation}\label{eq:visc}
\nu = \alpha c_s H,
\end{equation}
where $c_s$ is the sound speed and $\alpha$ is the dimensionless parameter between 0 and 1, describing the efficiency of angular momentum transport. Recent numerical simulations of accretion disks suggest that $\alpha \sim 0.01-0.1$ \citep[e.g.;][]{Hawley+1995,Stone+1996}. With the SS73 $\alpha$-disk model that many works still follow, we have parametrized our ignorance on the physical mechanisms responsible for the transport of angular momentum outward, which is required to enable accretion of material inward. The excess of energy will be used to heat the gas and is ultimately radiated away. The emerging flux at each radius, assuming the same accretion rate over the disk, is given by
\begin{equation}\label{eq:fdisk}
F_{d}(R) = \frac{3GM\dot M}{8\pi R^3} \left[ 1 - \left(\frac{3R_{s}}{R}\right)^{1/2}\right].
\end{equation}
From here, the effective temperature can easily be derived through the Stefan-Boltzmann relation $F = \sigma T_{eff}^4$ ($\sigma \simeq 5.67 \times 10^{-5}$\,erg\,cm$^{-2}$\,s$^{-1}$\,K$^{-4}$). Therefore, we see that $R$, $M$ and $\dot M$ determine the structure of the disk in the standard model. Also, SS73 found that for fixed values of $M$ and $\dot M$, the disk can be divided into three regions depending on $R$, that determine the gas density profile: the {\it outer region} at a large $R$, in which gas pressure dominates radiation pressure and the opacity is controlled by free-free absorption; the {\it middle region} at smaller $R$, in which gas pressure is dominated by radiation pressure but the opacity is mainly due to electron scattering; and the {\it inner region} at very small $R$, where radiation pressure dominates over the gas pressure, but still scattering dominates the absorption opacity. For some values of the accretion rate, the middle region may not exist. For a $10^6$\msun black hole accreting at 1\% of the Eddington limit, and assuming $\alpha=0.1$ the boundary between the inner and middle regions is at $R_{im} \sim 40 R_g$, and between the middle and outer regions at $R_{mo}\sim 700 R_g$.

Since the focus of this review is on the X-ray band, where emission originates at the smallest scales, we summarize here the important equations of SS73 for the radiation-dominated case (inner region):

Half-thickness
\begin{equation}
z \mathrm{[cm]} = 3.2\times 10^6 m \dot m \left(1-r^{-1/2}\right)
\end{equation}

Surface density
\begin{equation}\label{eq:u0}
    u \mathrm{[gr\,cm^{-2}]}= 4.6 \alpha^{-1} \dot m^{-1} r^{3/2} \left(1-r^{-1/2}\right)
\end{equation}

Midplane temperature
\begin{equation}\label{eq:tmid}
T \mathrm{[K]} = 2.3\times10^7 \left(\alpha m\right)^{-1/4} r^{-3/8}
\end{equation}

Optical depth\footnote{Note that this expression differs from the one appearing in SS73, in the  radial dependency. In SS73 it appears as $\propto r^{-93/32}$, which is likely incorrect, as it would imply an optically-thin disk. Equation~\ref{eq:tau} was derived using Eqs.~\ref{eq:u0}, \ref{eq:tmid} and \ref{eq:n} into  $\tau=\sqrt{\sigma_T\sigma_{ff}}u_0$, with $\sigma_T=0.4$ and $\sigma_{ff} = 0.11 T^{-7/2}n$, both in cm$^2$/gr.
}

\begin{equation}\label{eq:tau}
\tau = 8.2\times10^{-5} \alpha^{-17/16} m^{-1/16} \dot m^{-2} r^{57/16} \left(1-r^{-1/2}\right)^{-2}
\end{equation}

Here we used the dimensionless parameter definitions
\begin{equation}
m \equiv \frac{M}{M_{\odot}} \ \  \ \ \dot m \equiv \frac{\dot M}{\dot M_{Edd}}
\ \  \ \  r \equiv \frac{R}{3R_s}
\end{equation}
Then we can also calculate the scale height $H = z/r$, and the density as
\begin{equation}\label{eq:n}
n \mathrm{[cm^{-3}]}= \frac{u}{2m_p z}.
\end{equation}
So we also have an estimate of the pressure of the gas.

The SS73 $\alpha$-disk model has been the foundation of subsequent research on accretion disk theory, as it provides a very simple approximation to the global disk properties. However, this prescription does not adequately explain many observational aspects of AGN, such as their UV luminosity and variability (e.g. \citealt{Alloin+1986} and see Section~\ref{sec:timing}). Furthermore, the SS73 model is both thermally and viscously unstable, and thus it cannot be a correct solution for the inner regions of the accretion disk\footnote{Models that do not have these instabilities include magnetically supported models (e.g., \citealt{begelman2007}), which in general have rather different physical conditions as a result.
}. Theoretical work and numerical simulations have identified MRI
\begin{marginnote}[]
    \entry{MRI}{
    {\it Magneto-rotational instability}; fundamental property of ionized accretion flows that generates turbulence, acting as effective viscosity to transport angular momentum outward.}
\end{marginnote}
to be the leading process for turbulence in the accretion disk and thus as a main source of pseudo-viscosity \citep{Balbus+1991}, suggesting that accretion disks are not the idealized steady-state equilibrium assumed by the SS73 model. Thus, the inner regions of luminous disks in AGN  (particularly the radiation-pressure-dominated regions) are still poorly understood theoretically, and it is possible that the interpretation of future observational data may lead to changes in current paradigms.

\subsection{Relevant Variability Timescales}\label{sec:timescales}

Accretion of matter onto a black hole is an inherently turbulent phenomena, which yields strong variability at many different time scales. It is thus relevant to review the characteristic times, as they provide reasonable estimates for the observed variability.  

Perhaps the most fundamental timescale is the {\it light crossing time} for photons to travel from the corona to the disk, and from the disk to the observer. In flat space-time (i.e., without relativistic corrections), the light-crossing time is simply $t_\mathrm{lc} = R/c$, where $R$ is the mean distance traveled by the observed photons. This is the timescale responsible for the reverberation time delays observed from IR to X-rays, due to the irradiation of different gas flows around the central engine (Section~\ref{sec:reverberation}).

The other different timescales related to variability in AGN can be divided into two main groups; those produced by {\it accretion processes}, which will then characterize the properties of the infalling material and its overall structure; and those produced by {\it microscopic processes}, which are connected with the local properties of the plasma (e.g., thermodynamics and ionization state), and with its composition (elemental abundances and atomic properties).

\begin{figure*}
\begin{center}
\includegraphics[width=\textwidth]{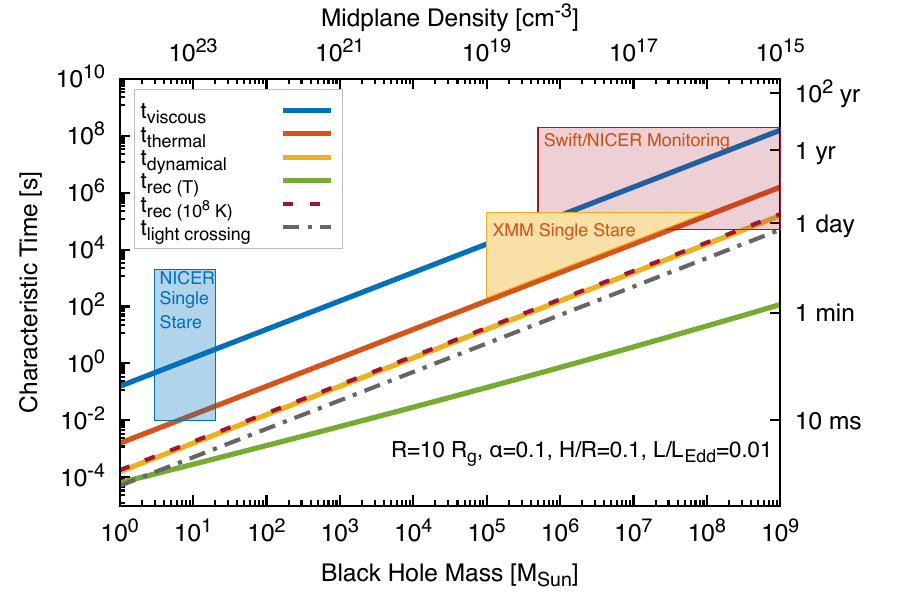}
\caption{Characteristic timescales relevant for an accretion flow in the innermost regions as a function of the black hole mass. The recombination time $t_\mathrm{rec}(T)$ is shown assuming the disk mid-plane density (Eq.~\ref{eq:tmid}), as well as the typical temperature ($T=10^8$\,K) of an X-ray irradiated disk surface. The shaded boxes indicate the timescales probed in typical {\xmm}, {\swift} and {\nicer} campaigns. (Note that {\nustar} and {\xrism} are capable long stares that probe timescales similar to {\xmm} for AGN, but because they both are in Low-Earth Orbit, they have orbital gaps every $\sim 90$ minutes that make traditional Fourier timing analyses more cumbersome, e.g., \citealt{lewin22}.)}
\label{fig:times}
\end{center}
\end{figure*}

For the accretion flow, the most relevant timescales are:

\begin{itemize}

\item Dynamical time:

\begin{equation}
    t_\mathrm{dyn} \sim \frac{1}{\Omega} = \frac{t_\mathrm{orb}}{2\pi},
\end{equation}
where $\Omega$ is the angular frequency $\Omega = (GM/R^3)^{1/2}$, and $t_\mathrm{orb}$ is the orbital time (i.e., the time it takes a particle at a distance $R$ to complete one revolution around the black hole). 
The angular frequency is sometimes interpreted as the cause of rare quasi-periodic signals seen in a few AGN and TDEs (\citealt{gierlinski08, pasham19, masterson25} and see Section~\ref{sec:QPE}).

\item Hydrostatic time:

\begin{equation}
    t_\mathrm{hydro} \sim \frac{H}{c_s},
\end{equation}
is the characteristic time under which the different pressures (e.g.; gas, radiation, gravity) present in the system cancel out such that hydrostatic equilibrium is achieved. This time is mainly controlled by the sound speed in the medium, as no mechanical changes to density or pressure can propagate faster than $c_s$. Using the fact the disk scale height is $H = c_s / \Omega$, then 

\begin{equation}
    t_\mathrm{hydro} \sim \frac{1}{\Omega} = t_\mathrm{dyn}.
\end{equation}

\item Thermal Time:

\begin{equation}
    t_\mathrm{th} \sim \frac{1}{\alpha\Omega} = \frac{1}{\alpha} t_\mathrm{dyn},
\end{equation}
is the timescale under which a disk annulus will cool down by releasing all its thermal energy if the heating mechanisms are instantaneously suppressed. It also indicates the shortest time under which the disk emission is expected to change as a result of variations in heating by viscosity.

\item Viscous time:

\begin{equation}
    t_\mathrm{visc} \sim \frac{R^2}{\nu},
\end{equation}
sometimes also referred to as the {\it radial infall timescale}, it is the time over which the viscosity $\nu$ in the flow will flatten any gradients in the surface density \citep{Armitage+2022}. Invoking the SS73 viscosity prescription (\ref{eq:visc}):

\begin{equation}
    t_\mathrm{visc} \sim \frac{1}{\alpha\Omega}\left( \frac{H}{R} \right)^{-2} = \frac{1}{\alpha}\left( \frac{H}{R} \right)^{-2} t_\mathrm{dyn},
\end{equation}

which in other words, is the characteristic time under which viscosity leads to transport of angular momentum outward allowing for material inflow. A more extensive discussion on these timescales can be found in reviews dedicated to the physics of accretion flows, such as \cite{Frank+2002} and \cite{Armitage+2022}.
\end{itemize}

\begin{textbox}[h]\section{Summary of Characteristic Time Scales}
The relevant timescales in the innermost regions close to the black hole can be summarized as follows:

\subsection{Light Crossing Time}
\begin{equation}
    t_\mathrm{lc} = R/c \sim 50\,[\mathrm{s}] \left(\frac{M}{10^6M_{\odot}}\right) \left(\frac{R}{10R_g}\right)
\end{equation}

\subsection{Accretion Flow}
\begin{equation}\label{eq:tdyn}
    t_\mathrm{dyn} = \frac{1}{\Omega} \sim 3\,[\mathrm{min}] \left(\frac{M}{10^6M_{\odot}}\right) \left(\frac{R}{10R_g}\right)^{3/2}
\end{equation}
\begin{equation}
    t_\mathrm{th} = \frac{1}{\alpha\Omega} \sim 27\,[\mathrm{min}] \left(\frac{0.1}{\alpha}\right) \left(\frac{M}{10^6M_{\odot}}\right) \left(\frac{R}{10R_g}\right)^{3/2}
\end{equation}
\begin{equation}
    t_\mathrm{visc} = \frac{1}{\alpha}\left(\frac{H}{R}\right)^{-2}\frac{1}{\Omega} \sim 44\,[\mathrm{hours}] \left(\frac{0.1}{\alpha}\right) \left(\frac{0.1}{H/R}\right)^2 \left(\frac{M}{10^6M_{\odot}}\right) \left(\frac{R}{10R_g}\right)^{3/2}
\end{equation}

\subsection{Microphysics}

\begin{equation}
    t_\mathrm{PI} \sim  3.1 \times 10^{-6}\,[\mathrm{ms}] \left( \frac{0.1}{\kappa_\mathrm{x}} \right) \left( \frac{0.1}{\lambda_\mathrm{Edd}} \right) \left(\frac{M}{10^6M_{\odot}}\right) \left(\frac{R}{10R_g}\right)^2
\end{equation}

\begin{equation}
    t_\mathrm{rec} \sim 7\,[\mathrm{min}] \left( \frac{n}{10^{15}} \right)^{-1}  \left( \frac{T}{10^8}\right)^{1.7}
\end{equation}
\end{textbox}

In terms of microscopic processes, there are timescales concerning the interaction of radiation and matter. In the context of accretion disks in AGN, where there is strong irradiation of X-rays from the hot corona, these are mostly related to the microscopic properties of the gas. In general, there is a long list of atomic processes that can occur at very different timescales, and can also differ substantially among ions with different nuclear and electric charges. The two most important timescale are the {\it photoionization} and {\it recombination} times, which are discussed more extensively in the Appendix. 
All these timescales are summarized in the Sidebar entitled {\it Summary of Characteristic Time Scale}.

The characteristic timescales discussed above are shown as a function of the black hole mass in Figure~\ref{fig:times}, compared to the duration of typical X-ray observing campaigns. The curves are computed for typical parameters in the innermost regions of bright AGN; e.g., inner radius ($10 R_g$), alpha parameter ($\alpha=0.1$), aspect ratio ($H/R=0.1$), and luminosity ($L=1\% L_\mathrm{Edd}$). The density is assumed to be that of the disk mid-plane in a SS73 model. In such a case, the recombination time is much shorter than any of the other relevant times, with the exception of the photoionization time, which is much shorter by several others of magnitude, and thus not shown in the figure (for simplicity, we have followed the approach of \cite{Garcia+2013}, assuming a pure hydrogen and constant density medium).

However, the recombination time is also a strong function of the gas temperature. In the case of a disk illuminated by the strong radiation from the X-ray corona, the temperature at the surface can reach much higher values, even close to the Compton temperature ($T\sim10^7-10^9$~K, independently of the black hole mass),
\begin{marginnote}[]
\entry{Compton Temperature}{
$T_\mathrm{C} = <E>/4k$, temperature where photons undergoing repeated electron scatterings reach balance in energy lost and gained ($<E>$ is mean photon energy).}
\end{marginnote}
as predicted by ionized reflection calculations \citep[e.g.;][]{garcia10,gar13a}. This hot layer can still have a considerable optical depth (e.g.; $\tau\sim1-3$), depending on the strength of the incident flux. At such high temperatures, the recombination times become much longer, even at the relatively high density predicted for the disk mid-plane. For example, for a temperature of $T=10^8$\,K, the recombination time coincidentally becomes almost identical to the dynamical time scale across all masses, at a radius of 10\,$R_g$. This situation is exacerbated if the surface density is lower than at the mid-plane, which is more plausible than the density being constant throughout the disk's vertical extension.

Other important times related to the microphysics of the accretion flow, such as the thermal equilibrium time and the photon propagation time are discussed elsewhere \citep[e.g.;][]{Garcia+2013,Rogantini+2022,Sadaula+2023}. These recent works represent the initial steps of a complex problem of time-dependent photoionization coupled with the macroscopic properties of the accreted material. Their results clearly indicate the need to expand these studies, to properly account for all the physics relevant to accreting systems.

\section{KEY X-RAY OBSERVABLES OF STANDARD AGN}\label{chap:AGN}

The process of growing a black hole through the accretion of matter releases an extraordinary amount of energy (Eq.~\ref{eq:eta}). For luminous AGN of $10^{6}-10^{8}$~\msun, most of that emission is released in the EUV and in X-rays from very close to the black hole (within $\sim 10 R_g$). {\em With the EUV obscured by Galactic photoelectric absorption, X-rays offer a powerful means to study these energetic environments.}. Thus, X-rays directly probe how black holes grow via accretion and how AGN are powered. 

\begin{figure*}
\begin{center}
\includegraphics[width=\textwidth]{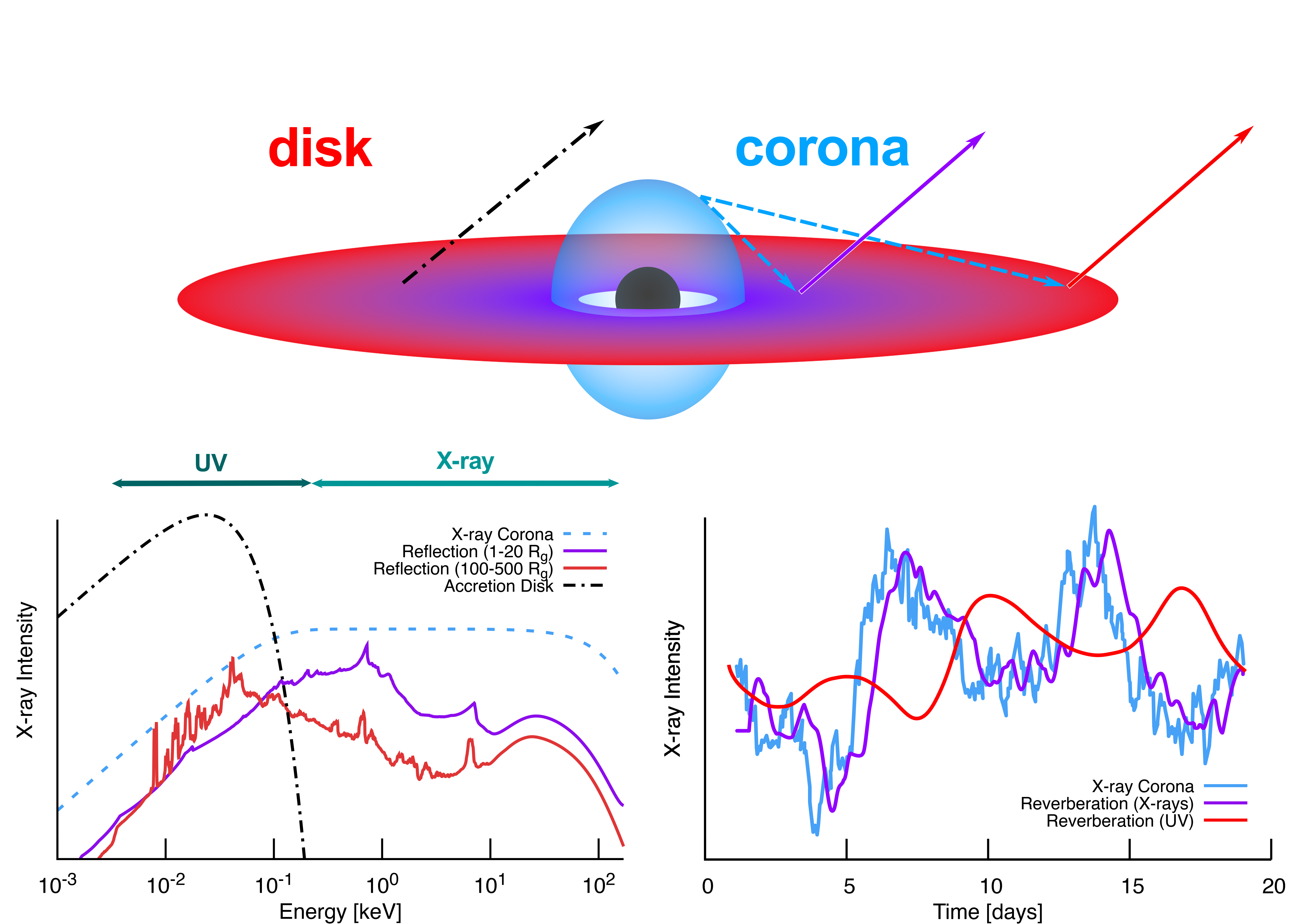}
\caption{Schematic view of the inner regions of an AGN (top), together with the corresponding UV-to-X-ray spectra and light curves. {\em Bottom-left:} The AGN continuum consists of thermal blackbody emission from the disk (black) and X-rays from the corona (blue). The corona will inevitably irradiate the disk, causing the reflection spectra, where the inner-disk reflection dominates in X-rays (purple) and outer-disk reflection peaks in the UV (red). {\em Bottom-right:} The reflection spectra are delayed with respect to the continuum (a process known as ``reverberation"). X-ray reverberation time delays in purple are short ($\sim$~minutes for a $10^{7}$\msun\ black hole, or tens of $R_g$), while UV reverberation time delays in red are longer ($\sim$days for a $10^{7}$\msun\ black hole, or hundreds of $R_g$). Figure schematics courtesy of Joanna Piotrowska (Caltech). X-ray light curve schematic courtesy of Guglielmo Mastroserio (INAF).  
}
\label{fig:sed}
\end{center}
\end{figure*}

Figure~\ref{fig:sed} schematically summarizes the topics in this Section. The primary emission of an AGN is seen from the disk (Sec.~\ref{sec:disk-corona}) and corona (Sec.~\ref{sec:corona}). The emission from the corona is polarized (Sec.~\ref{sec:polarization}), highly variable (Sec.~\ref{sec:timing}), and in addition to shining the observer, it also shines back on the accretion disk, causing reprocessed emission, often referred to as ``reflection" (Sec.~\ref{sec:reflection}). We see this reflected light in both X-ray fluorescence lines and in the UV/optical, and it is delayed with respect to the coronal continuum because of the additional light-travel path it has taken (Sec.~\ref{sec:reverberation}). The corona also shines light on more distant gas flows around the black hole, from the broad line region and dusty torus (Sec.~\ref{sec:blr-torus}) to outflows that may be responsible for AGN feedback (Sec.~\ref{sec:outflows}). Throughout this section, we often compare the results of AGN to black hole X-ray binaries (see Sidebar entitled {\it Stellar-Mass Black Holes}).

\begin{textbox}[t]\section{STELLAR-MASS BLACK HOLES}

Stellar-mass black holes with stellar companions are intriguing and convenient systems to test our theories of accretion. Whereas AGN fueling from the host galaxy is notoriously complex, in black hole X-ray binaries (BH XRBs) with low-mass companions, the feeding is well understood to be from the companion star filling its Roche lobe. As such, one observes an accretion disk, corona and jets, but no torus or broad line region. Whereas AGN emit across the entire electromagnetic spectrum, in BH XRBs, the accretion disks are hotter, and so both the disk and corona emit in X-rays. This allows us to track the disk-corona connection with a single X-ray instrument. Finally, the variability scales with the mass of the black hole (Fig.~1). In BH XRBs the typical light travel time of reverberation out to $10~R_g$ is $\sim 0.1$~ms, and therefore, even in a typical {\em NICER} observation of just a few kiloseconds in duration, we can measure (with better signal-to-noise) the kinds of time lags it would take days to even {\em years} to measure in AGN.

The scaling between AGN and stellar-mass black holes is demonstrated in Fig.~\ref{fig:lags}, where we show an example of the low- and high-frequency X-ray lags for a bright variable AGN Ark~564 ($M_{\mathrm{BH}}\sim 10^{6}$\msun) compared to the black hole low-mass X-ray binary MAXI~J1820+070 ($M_{\mathrm{BH}}\sim 10$\msun) taken during the luminous hard state, when $L/L_{Edd} \sim 0.1$. They look remarkably similar, but the amplitude of the lag is $10^{5}$ times shorter for the stellar-mass black hole. (see Sec.~{\ref{sec:reverberation}} for more details on reverberation).

This scaling with mass affords the opportunity to measure reverberation lags at different stages of accretion in BH XRBs. While the typical duty cycle of an AGN is expected to be $\sim 10^{6-8}$~years, the equivalent in an X-ray binary is $\sim 1$ year. Indeed, the typical duration of an outburst from and X-ray binary is several months. By measuring X-ray reverberation lags at several epochs, we can quantify how the corona, disk and jet evolve together over time \citep{demarco17,Kara19,Wang2021}. 
\end{textbox}


\subsection{Spectral Properties of the X-ray Continuum}\label{sec:spectral}

\begin{textbox}[h]\section{Origin of the Power-law Continuum}

\subsection{Inverse Compton Scattering}

The ``standard" Compton scattering process considers the collision of an energetic photon with an electron at rest. The photon transfers some of its initial energy $E_i$ to the electron's kinetic energy, emerging after the interaction with a final energy $E_f$ given by the well known Compton formula:
\begin{equation}
    E_f = \frac{E_i}{1 + E_i/m_ec^2 (1-\cos{\theta})},
\end{equation}
where $\theta$ is the angle between the photon's initial and final trajectory. However, if the electron has a kinetic energy comparable to the energy of the photon, then energy can be transfer in the opposite way---from the electron to the photon---a process known as {\it inverse Compton scattering}, which can be particularly important for a gas with relativistic electrons. Considering the initial and final energies and scattering angles in both the laboratory and electron's rest frames, an equivalent Compton formula for the energy shift can be written, but its derivation is much too long for this review (details can be found in specialized literature such as \citealt{Rybicki79}). However, in the limit when the photon energy is much less than the electron's rest energy ($E_i' \ll m_ec^2$, in the electron's rest frame), for typical collisions (initial and final angles $\sim \pi/2$), the observed photon's final energy is 
\begin{equation}
    E_f \approx \gamma^2 E_i,
\end{equation}
where $\gamma = 1/\sqrt{1-v^2/c^2}$ is the Lorentz factor. Since $\gamma$ is always larger than unity for relativistic electrons, photons can gain energy by factors of several after a single scattering.

\subsection{Thermal Comptonization}

Inverse Compton scattering can take place when the UV or soft X-ray photons from the thermal disk undergo multiple scatterings in a hot ($\sim 10^9$\,K) and optically thin ($\tau\sim1-3$) X-ray corona, which explains the power-law X-ray continuum observed from accreting black holes (Sec.~\ref{sec:corona}). The cutoff at high energies is related to the electron temperature of the plasma.

\subsection{Bulk Motion Comptonization}

If electrons are accelerated to nearly the speed of light by non-thermal processes (such as in the accretion flow inside the ISCO radius, or through strong magnetic fields) and have trans-relativistic bulk motions, they can still upscatter photons to a hard non-thermal distribution, even if the electrons are cold \citep{Blandford1981,Chakrabarti1995}. In this case, the plasma can have a lower temperature and lower optical depth than in the thermal Comptonization picture.
\end{textbox}

\subsubsection{The accretion disk and its relationship to the X-ray corona}\label{sec:disk-corona}

Fundamentally, as matter funnels towards the black hole through the accretion disk, angular momentum is lost via viscous processes, heating up the disk, and producing multi-temperature thermal emission. This emission is seen directly in the optical/UV and often referred to in undergraduate astrophysics textbooks as ``the big blue bump". For luminous black holes of $10^{6}-10^{8}$~\msun, most of the power output from the thermal disk peaks in the EUV, in the unobservable gap imposed by Galactic photoelectric absorption. This complicates studies attempting to calculate the bolometric luminosity, $L_\mathrm{Bol}$,
\begin{marginnote}[]
    \entry{Bolometric luminosity, $L_\mathrm{Bol}$ } {
    Total intrinsic power released from the accretion process across all wavelengths.
    }
\end{marginnote}
though there have been several attempts to bridge that gap, either by developing physical models for the accretion flow \citep[e.g.;][]{marconi04,done12}, or by assuming agnostic phenomenological models \citep[e.g.;][]{grupe10}. As such, caution is warranted in reading estimates of bolometric luminosity, and even more so the Eddington ratio, $\lambda_\mathrm{Edd}\equiv L_\mathrm{Bol}/L_\mathrm{Edd}$, used as a proxy for the mass accretion rate, because that also folds in uncertainties in our measurements of the black hole mass. 

The accretion process also leads to strong X-ray emission, as thermal disk photons (so-called ``seed" photons) interact with hot ($\sim10^9$\,K) electrons in the central corona, and are inverse Compton upscattered to X-ray energies (Sec.~\ref{sec:corona}). Several studies have attempted to calculate the ratio of X-ray to bolometric luminosity: Some focus on small samples in the nearby Seyferts \citep{vasudevan07,jin12,duras2020}, in which the broadband SED can be measured and modeled with care (also accounting for the fact that AGN vary and thus optical and X-ray measurements should be made simultaneously). Others have corroborated these studies with large surveys, such as COSMOS \citep{lusso2010} and eROSITA eFEDS \citep{liu2022}, where the sheer number of AGN precludes careful studies of any one, but improves statistics. {\em From these varied approaches, there is general agreement that about $1-20\%$ of the bolometric luminosity is radiated in X-rays}, though this value depends strongly on luminosity and mass accretion rate: The ratio of X-ray-to-bolometric luminosity\footnote{There have been many studies working to quantify how much energy is released in the accretion disk vs. the corona. Often one sees this parametrized as $\alpha_\mathrm{ox}$ \citep{Tananbaum1979}, which is the power-law slope between the $2500$\,\AA\ luminosity and the 2\,keV luminosity. This parametrization has the benefit of being model independent, and easy to measure. A smaller $\alpha_\mathrm{ox}$ generally means a lower ratio of X-ray-to-bolometric luminosity. The corona-to-disk power is also often quantified as a bolometric correction or `K-correction', where $\kappa_\mathrm{x}=L_\mathrm{Bol}/L_\mathrm{x}$. There is a general correlation between $\alpha_\mathrm{ox}$ and $\kappa_\mathrm{x}$. While model-dependent, we choose to quote the ratio of X-ray to bolometric luminosity ($L_\mathrm{x}/L_\mathrm{Bol}$) because we find it most intuitive.} decreases from $\sim 10$\% in AGN with $L_\mathrm{Bol}\sim 4 \times 10^{42-44}$~erg/cm$^{2}$/s, down to $\sim 1\%$ in AGN with higher luminosities, $L_\mathrm{Bol}\sim 4\times 10^{46}$ erg/cm$^{2}$/s. Moreover, there is a noted dependence on mass accretion rate, where the {\em X-ray-to-bolometric luminosity decreases with increasing \mdot}: from $\sim 10\%$ at $\lambda_\mathrm{Edd}\sim0.01$ down to $2\%$ at $\lambda_\mathrm{Edd}\sim1$. Interestingly, these numbers so far do not appear to evolve with redshift, at least up to $z=3.5$ (though note the interesting new results on high-redshift ``Little Red Dots" with {\it JWST} that, if truly all AGN, appear to be very X-ray weak; \citealt{yue2024,maiolino2024}).

\subsubsection{The ubiquitous hard X-ray corona: Phenomenology}\label{sec:corona}

Coronal X-ray emission provides crucial insights into the energetic processes near supermassive black holes, as it probes the innermost regions and thus it is key to understanding AGN emission, black hole feedback, and galaxy evolution. Observationally, the X-ray continuum from most Seyfert AGN is described by a power-law with photon index $\Gamma$ and an exponential cutoff at high energies ($E_\mathrm{cut}\sim300$\,keV). 
Physically, this is interpreted as the result of soft UV disk photons entering a hot ($T\sim10^{9}$\,K) and optically-thin ($\tau\sim 1-3$) corona. Low-energy photons gain energy through multiple inverse Compton scatterings with hot electrons, producing a power-law continuum, in a process commonly known as {\it Thermal Comptonization} (see Sidebar entitled {\it Origin of the Power-law Continuum}).
The energy gained by the coronal photons is limited by the mean kinetic energy of the relativistic electrons, and thus the power law falls off abruptly at a cutoff energy ($E_\mathrm{cut}$) of a few times the electron temperature of the plasma \citep{Titarchuk+1995}.

\begin{figure}
\centering
\includegraphics[width=\textwidth]{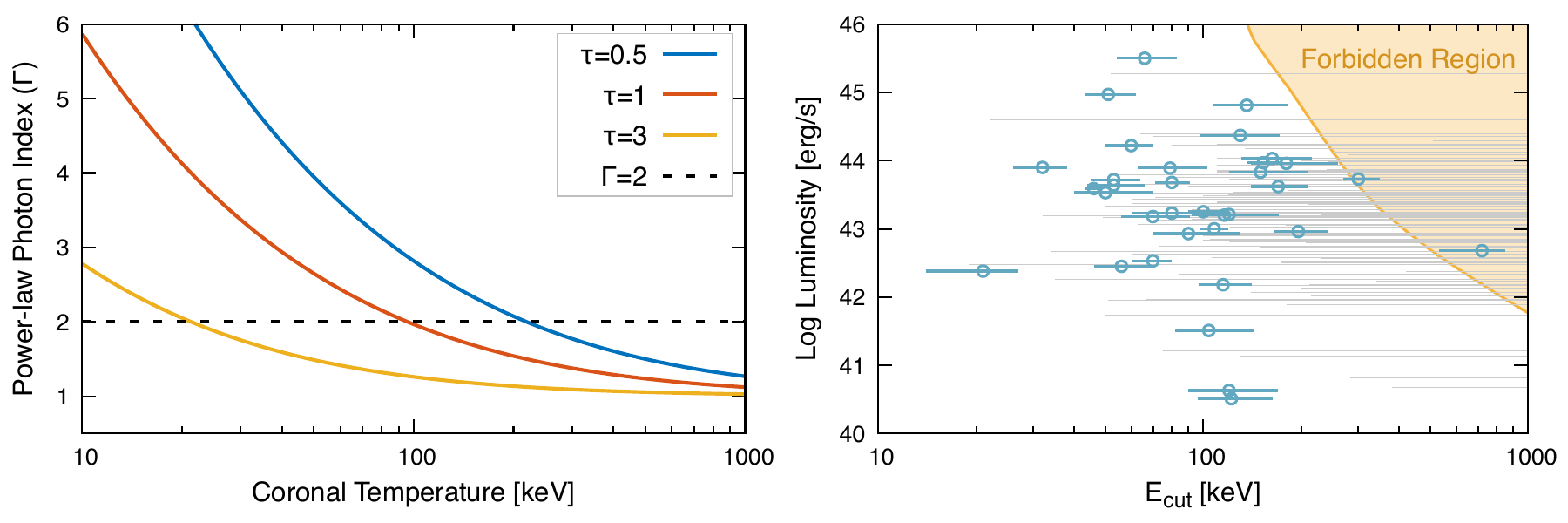}
\caption{{\it Left}: Photon index $\Gamma$ as a function of coronal temperature for constant  optical depths computed with the relationship from Eq.~\ref{eq:gamma1}. If the slope of the continuum (i.e.; $\Gamma=2$, black-dashed line), and the electron temperature is derived from the cutoff energy (Eq.~\ref{eq:ecut}), one can easily estimate the optical depth of the corona. {\it Right}: Compilation of measured coronal temperatures and luminosities for local AGN from \cite{Bertola+2022}. Formal constraints are shown with blue-empty circles, while lower-limits are shown as thin-gray bars. The pair-production forbidden region assumes a slab geometry in a $10^8$\msun\ black hole \citep[different geometries and masses define slightly different regions, for more details see,][]{Kammoun+2024}.
}
\label{fig:corona}
\end{figure}

Observationally, it is relatively straightforward to measure the slope of the continuum ($\Gamma$) and its cutoff at high energies ($E_\mathrm{cut}$). With these quantities, one can derive the electron temperature ($T_e$) and optical depth ($\tau$) of the corona by implementing the empirical relationships:
\footnote{Eq.~\ref{eq:gamma1} is appropriate for most cases, sources with faint and/or soft non-thermal emission might indicate emission from a corona with much lower optical depth ($\tau<1$), in which case an alternative expression should be used, e.g.,
\begin{equation}\label{eq:gamma2}
    \Gamma = 1 + \frac{-\ln{\tau} + 2/(3+\theta_e)}{\ln{12\theta^2_e}+25\theta_e},
\end{equation}
\citep{Sridhar+2020}.
}
\begin{equation}\label{eq:ecut}
    E_\mathrm{cut} = (2-3)kT_e,
\end{equation}
where the exact value of the proportionality constant depends on the assumed coronal geometry \citep{Petrucci+2001}, and
\begin{equation}\label{eq:gamma1}
    \Gamma = -\frac{1}{2}+\sqrt{\frac{9}{4} + \frac{1}{\theta_e\tau(1+\tau/3)}} : \tau \gtrsim 1,
\end{equation}
where $\theta_e=kT_e/m_ec^2$, and $m_ec^2=511$\,keV is the electron's rest mass energy \citep{Pozdnyakov+1983,Lightman+1987,Haardt+1993}. From these relationships, it is clear that objects with $\Gamma\sim 2$ are unlikely to have coronal temperatures above $\sim 100$\,keV, which implies observed cutoffs of $E_\mathrm{cut}\sim 300$\,keV (Figure~\ref{fig:corona}, left).


Systematic studies with the {\swift}/BAT sample suggest higher accretion rate AGN exhibit steeper photon indices and cooler coronae \citep[e.g.;][]{Ricci+2018, hinkle2021}, but follow-ups with smaller \nustar\ samples have not confirmed these correlations \citep[e.g.;][]{Tortosa+2018,Kamraj+2022,Serafinelli+2024}. This discrepancy may stem from \nustar's narrower energy range (3–79~keV) and smaller sample sizes, though its higher signal-to-noise allows better parameter estimation by disentangling degeneracies between the continuum slope ($\Gamma$) and high-energy cutoff ($E_\mathrm{cut}$). Correlations between coronal parameters and accretion rate reflect the complex interplay between heating and cooling mechanisms. For example, higher accretion rates should increase UV photons cooling the corona, steepening $\Gamma$ and lowering $E_\mathrm{cut}$ \citep{kara2017}. A lack of correlation between $\Gamma$ and $E_\mathrm{cut}$ with accretion rate could point towards a more complex physical scenario beyond the simple thermal Comptonization picture.

In fact, an important caveat to the present discussion is that historically only thermal Comptonization has been assumed as the physical origin of the corona. However, as we explore in the next Section, new research on relativistic plasma physics is rapidly changing this picture, pointing towards {\it Bulk Motion Comptonization} as an alternative scenario (see Sidebar entitled {\it Origin of the Power-law Continuum}). This could dramatically change the physical inferences derived from X-ray observations (e.g.; with temperature really being driven by the kinetic energy of bulk motion in some cases).

\subsubsection{The ubiquitous hard X-ray corona: Its origin and geometry}\label{sec:corona-origin}

The physical origin of the X-ray corona remains an active area of research. The presence of compact radio emission in most AGN \citep{Smith20} has led to suggestions that the magnetic fields responsible for jet emission could also confine and heat the corona \citep{behar18,king17}. Such a correlation between the corona and the jet is obvious in black hole X-ray binaries, where persistent radio jets are only observed in ``hard" accretion states, during which the coronal continuum dominates the X-ray spectrum \citep{fender04,markoff2005}. Other models do not require the corona to be the jet base. For example, 3D radiation MHD simulations of sub-Eddington disks by \cite{Jiang+2019} show a fairly hot ($\sim10^8$~K), optically-thin and spatially compact (within $10 R_g$) plasma that resembles the X-ray corona. More recently, \cite{Bambic+2024} demonstrated through shearing box simulations that vertical magnetic fields could efficiently power coronae in bright AGN and quasars without necessarily launching a strong jet. For further details on the role of magnetic fields in jets, disks, and coronae, see \citet{Davis+2020} and references therein.

Constraining the geometry of the X-ray corona, such as its size and location, is vital for testing different corona formation models. For instance, the original ``two-phase" model predicts a corona extending over the disk \citep{Haardt+1991}, while the jet-base model suggests a compact corona, confined to within $\sim 10 R_g$ \citep{Ghisellini+2004}. Microlensing and reverberation studies of quasars (Sec.~\ref{sec:reverberation}) have offered promising constraints on the size of the corona, indicating that it is centrally located and is confined to within tens of $R_g$, and is more compact than the optically-emitting accretion disk \citep[e.g.;][]{Pooley+2007,demarco13,Mosquera+2013,uttley14}. This has led to the adoption of the lamp-post geometry, where the X-ray corona is approximated by a point source above the accretion disk, potentially arising from shocks in an ejection flow or the base of a jet \citep[e.g.;][]{Matt+1991,Martocchia+1996,Miniutti+2004}. More realistic 3D coronal geometries are now being considered in recent research \citep[e.g.;][]{Wilkins+2016,Gonzalez+2017,Zhang+2019}.

Although the lamppost geometry is highly idealized, it allows for a more simple calculation of the illumination pattern of the accretion disk by using photon ray-tracing codes in full general relativity. The predicted illumination profile is in turn incorporated in X-ray disk reflection and reverberation models which are widely used to constrain properties of the inner disk and estimate the black hole spin (see Sec.~\ref{sec:reflection} on reflection and spin measurements, and Sec.~\ref{sec:reverberation} on reverberation). So even though the exact origin of the corona is not settled, we can use the corona to shine a light on the accretion disk closest to the event horizon.

\subsubsection{The ubiquitous hard X-ray corona: The microphysics and pair thermostat}

As discussed in the previous section, the exact physical origin of the X-ray corona is still a matter of research. More consensus exist on the microphysics of the corona. The temperature of the plasma is set by the balance of heating and cooling. Heating has traditionally been attributed to magnetic reconnection \citep[e.g.;][]{Galeev+1979,Merloni+2001,Sironi+2020}. However, if this were the only heating mechanism, X-ray coronae would be short-lived because Compton cooling has a much shorter timescale \citep{Beloborodov+2017}. Thus, recent works on particle-in-cell (PIC) simulations point to a more complex scenario, in which relativistic bulk motions of a colder plasma can efficiently Comptonize seed photons to high energies \citep[e.g.;][]{Sridhar+2021,Sridhar+2023,Gupta+2024}. These simulations also show that turbulence in a highly magnetized plasma can explain state transitions in stellar-mass black hole systems, when all relevant quantum electrodynamical processes are considered \citep{Nattila+2024}.

Compton cooling is driven by the number of high-energy photons escaping the corona, which gained energy via inverse Compton scattering\footnote{In general, electron scattering is a very ``socialist" process; the particle with more energy shares it with the less energetic one. Thus, photons will gain or lose energy depending on their relation to the kinetic energy of the electrons.}.
Even a small number of such photons produce copious amounts of electron-positron pairs to share the available thermal energy, effectively cooling the electrons until a balance is reached. This is known as the {\it pair-production thermostat model}. The maximum temperature allowed is directly related to the total X-ray luminosity of the corona. The inverse Compton cooling power is given by
\begin{equation}
    P_\mathrm{IC} = \frac{4}{3}\sigma_\mathrm{T}c\beta^2\gamma^2U_\mathrm{rad},
\end{equation}
where $\sigma_\mathrm{T}$ is the Thomson cross section, $\beta$ is the velocity of the electrons (in units of $c$), $\gamma$ is the Lorentz factor, and $U_\mathrm{rad}$ is the energy density of the radiation field incident to the corona. Thus, brighter disks will tend to cool the corona more efficiently\footnote{Formally, an effective cooling requires $U_\mathrm{rad}$ to be large, and thus the important quantity is the number of photons per unit volume.}, possibly faster than the mechanism responsible for the heating. In such a situation, the plasma will be out of thermal equilibrium, which leads to an energy imbalance, effectively quenching the coronal emission. This could very well be the reason why the coronal continuum in black hole binaries quickly becomes much softer and fainter when sources transition from the hard to the soft state \citep[e.g.;][]{Sridhar+2020}. However, it is unclear whether AGN undergo similar state transitions (just over very long time scales), or whether such a behavior is exclusive to stellar-mass black hole systems.

The right panel in Figure~\ref{fig:corona} shows the most recent coronal temperature estimates and their X-ray luminosity ($2-10$\,keV) from the compilation of \cite{Bertola+2022}, which follows the work of \cite{Fabian15}. Many of these measurements are lower limits, owing to limitations in current detectors. As most temperatures are in the range of $10^8-10^9$\,K, this requires the detection of cutoff energies of hundreds of keV, while current hard X-ray telescopes like \nustar\ are only sensitive up to $\sim 80$\,keV. However, these measurements generally agree well with the predictions from the pair-thermostat model. Crucially, none of the well determined temperatures are observed to violate the theoretical limit (\citealt{Fabian15}, and see \citealt{Kammoun+2024} for a recent review).

Another important aspect to note from Fig.~\ref{fig:corona} is that we do not observe a conglomeration of points at the thermostat limit, which is expected if the plasma is assumed to be composed of electrons following a purely thermal distribution of energy. Instead, several AGN are reported at much lower electron temperatures \citep[e.g.;][]{Balokovic+2015,Ursini+2016,kara2017,Tortosa+2018,Reeves+2021}. This spread of coronal temperatures may be indicative of bulk motion Comptonization or of a plasma with a mixture of thermal and non-thermal particles (a so-called {\it hybrid plasma}), as suggested by several authors \citep[e.g.;][]{Ghisellini+1993,Zdziarski+1993,Fabian+2017}. Understanding whether the corona is indeed a hybrid plasma has consequences for how it is powered through magnetic dissipation processes (i.e., via magnetic reconnection, instabilities, and turbulence; \citealt{Spitkovsky+2005, Comisso+2018, Sironi+2014}). Thus, {\it the observational characterization of X-ray coronae and their properties probes fundamental plasma physics.}

\subsection{Polarization Properties of the X-ray Continuum}\label{sec:polarization}

\begin{figure*}
\begin{center}
\includegraphics[width=\textwidth, trim={0.5cm 1cm 2cm 1cm}]{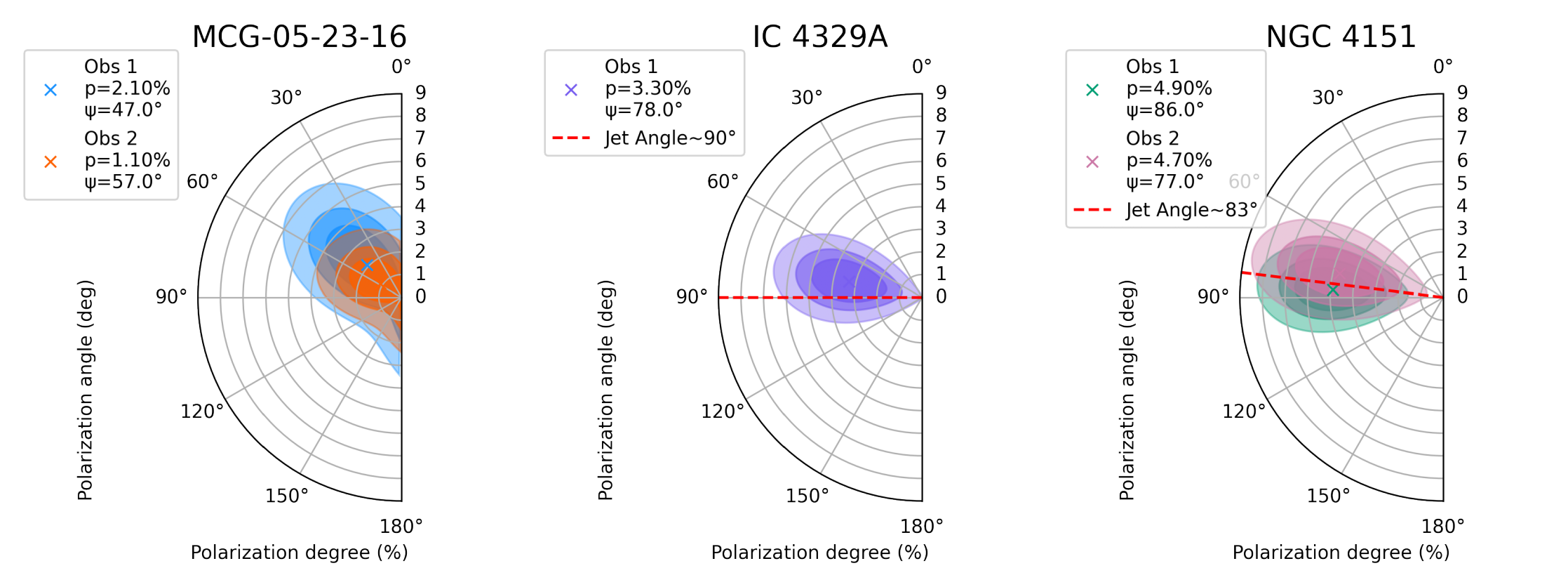}
\caption{X-ray polarization measurements for the three Type 1 radio-quiet AGN observed to date by {\ixpe}. For each source, a protractor plot shows the polarization degree and polarization angle in the $2-8$\,keV band using contours at 1, 2, and 3-$\sigma$ confidence levels. Individual observations for each source are shown with different colors. The orientation of a radio jet is shown for the two sources with reported estimates (red-dashed line). Original version of the Figure courtesy of Swati Ravi (MIT).
}
\label{fig:ixpe}
\end{center}
\end{figure*}

X-ray polarization has recently made an appearance in the high-energy astrophysics community with the launch of the {\it Imaging X-ray Polarimetry Explorer} \citep[\ixpe;][]{Weisskopf+2022} in December 2021. In general, linear polarization of light---in any energy band---measures the direction of the electric field vector, such that the degree of polarization (PD; often also denoted as $\Pi$) indicates the fraction of the intensity emitted in a preferential direction, while the polarization angle (PA; or $\Psi$) indicates what that direction is. 

Polarimetry observations of accreting black holes offer the opportunity to learn more about the geometry and morphological configuration of the regions emitting in X-rays. For radio-loud AGN, strong polarization is observed from the synchrotron jet emission. Conversely, radio-quiet AGN are environments relatively clean from synchrotron radiation and thus their polarization signal, albeit fainter, allows the study of accretion processes. In unobscured, Type~1 AGN, most of the polarization is likely taking place in the X-ray corona, driven by electron scattering, and thus measuring the PD and PA provides direct constraints on its geometry. In obscured AGN, polarization measurements can be used to describe the orientation of the obscuring torus \citep{ursini2023}.

Theoretical models predict several different possible geometries for the X-ray corona---including configurations such as spherical, conical, slab, or patchy---which will produce different polarization signals. In particular, the PA indicates the orientation of the corona, while the PD, and its dependence on energy, can discriminate between different model flavors \citep[e.g.;][]{Poutanen+1996,Veledina+2022}. For example, the relativistic outflowing plasma model proposed by \cite{Poutanen+2023} predicts a different dependency of the PD with energy for varying velocities of the scattering medium.

As of the writing of this Review, \ixpe\ has conducted observations of three Type 1 AGN: MCG$-$05-23-16 \citep{Marinucci+2022,Tagliacozzo+2023}, IC~4329A \citep{Ingram+2023}, and NGC~4151 \citep{Gianolli+2023}, though NGC~4151 is formally a Type 1.9, i.e. with significant line-of-sight obscuration. Their measured PA and PD are shown in the protractor plots in Figure~\ref{fig:ixpe} for different confidence levels. 
A significant PD is seen only in NGC~4151 at 1-8\%, which indicates scattering in the corona in this system. The PA of NGC~4151 and IC~4329A appear aligned with the jet (though the PD of IC~4329A is consistent with zero at the $3\sigma$ confidence level). A significant PA aligned with the jet would suggest that the corona is extended along the disk plane, and perpendicular to the jet. While the AGN measurements remain tentative, similar measurements of alignment of the PA and jet have been seen in BH X-ray binaries, such as Cyg~X-1 \citep{Krawczynski+2022} and GX~339-4 \citep{Mastroserio2024}.

At face value, the \ixpe\ measurements of Type~1 AGN challenge the results from microlensing and reverberation, which show the X-ray corona is more compact than the disk (Section~\ref{sec:corona}), and also challenge the simplest lamppost geometry for the corona, widely used in X-ray reflection measurements (Section~\ref{sec:reflection}). It appears that, in addition to {\em deeper} \ixpe\ observations of a larger sample of {\em truly unobscured AGN}, we require self-consistent modeling of spectra, timing and polarization with more sophisticated treatments of the corona (e.g., including also some radial extent of the corona, or a bulk outflow velocity). At this point, it is thus safe to argue that further observational and theoretical efforts are needed to reconcile these predictions.

More details on these measurements and those on Type 2 AGN can be found in the recent review by \citet{Marin+2024}, and references therein. A similar review for black hole X-ray binaries is given by \citet{Dovciak2024}, and a discussion comparing polarimetry measurements in both types of objects is provided by \cite{Saade+2024}.

\subsection{Variability Properties of the X-ray Continuum}\label{sec:timing}

Rapid variability is one of the hallmark features of an AGN, and, as discussed in Section~\ref{sec:intro}, allowed for the early realization that these enormous luminosities were powered by accretion on to a central supermassive black hole \citep{Lynden-Bell1969}.

\begin{figure*}
\begin{center}
\includegraphics[width=\textwidth]{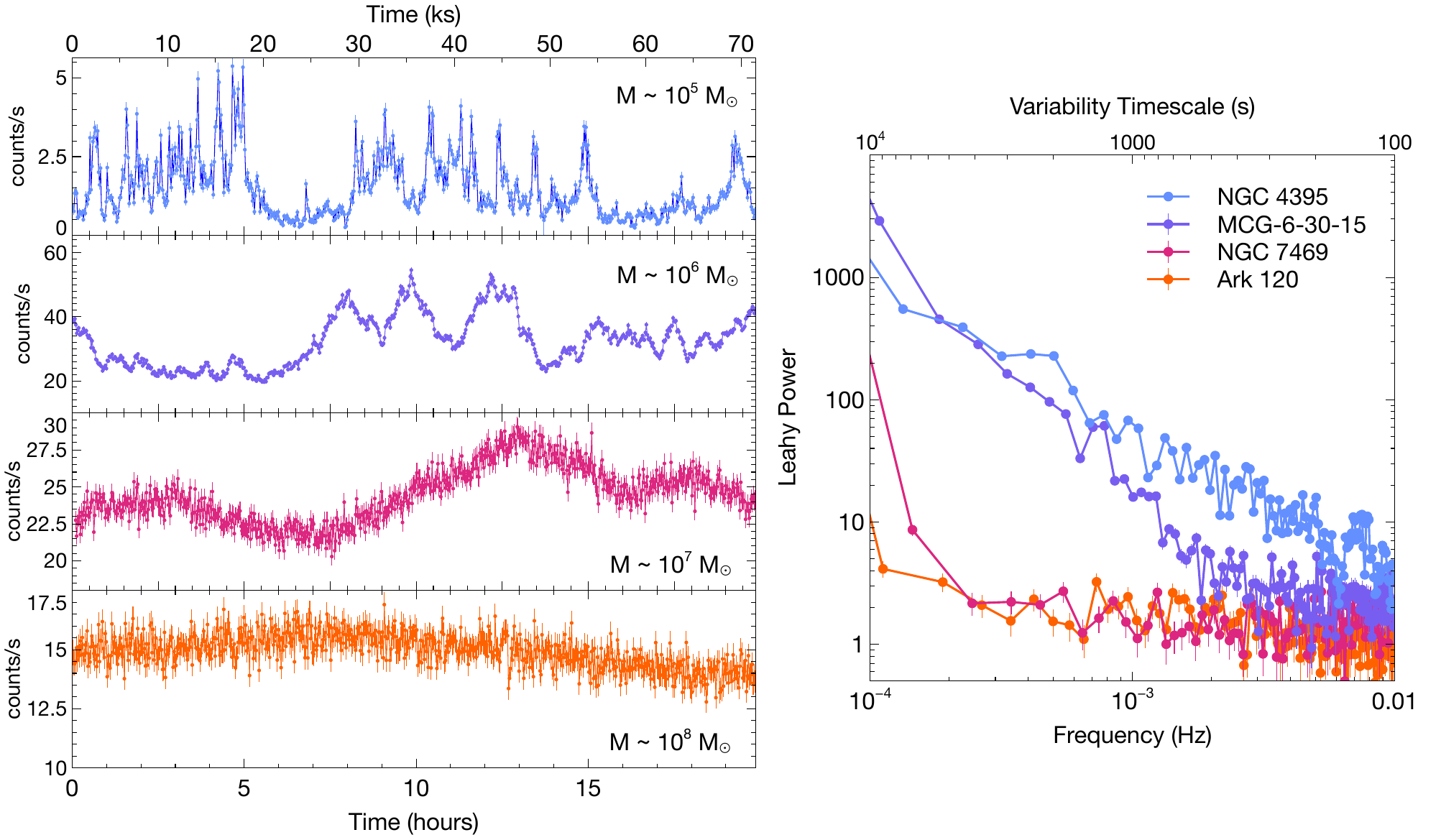}
\caption{{\em Left:} A comparison of the X-ray light curves of four representative AGN with masses of $10^{5}$\msun (top) to $10^{8}$\msun, measured with {\xmm} on timescales of hours. AGN vary stochastically, with smaller black holes varying more rapidly than larger. As such, the X-ray variance over a particular timescale is a strong indicator of the black hole mass \citep{Ponti2012}. {\em Right:} The corresponding PSDs for the same AGN. In the high frequency regime probed by {\xmm} ($\nu \sim [10^{-5}-10^{-2}$\,Hz]), AGN show a red noise power spectrum ($\alpha \sim 2$) that becomes dominated by Poisson noise at high frequencies. Lower mass black holes show power up to higher frequencies. For AGN that are roughly $10^{6}$\msun, we observe variability power above the Poisson noise limit on timescales roughly close to the thermal and viscous timescales (see Fig.~\ref{fig:times}).}
\label{fig:massPSD}
\end{center}
\end{figure*}

AGN timing analysis is difficult because the observed variability is stochastic (or random). This is demonstrated for four example AGN light curves in Fig.~\ref{fig:massPSD}, all observed with the {\xmm} observatory for 20~hours (or 70~ks in X-ray parlance). Unlike deterministic analyses, like the modeling of energy spectra from a non-variable sources or the timing of a strictly periodic sources, where the analysis is repeatable within the limits of measurement noise, AGN variability is unpredictable, and the signal itself is a noisy process. The ups and downs of an individual AGN light curve are not particularly meaningful: they are simply a random realization of an inherent underlying physical process. Therefore, for such non-deterministic data, statistical handling is required. One gains physical insights by determining the average, underlying variability properties (see \citealt{vaughan03} for a beautiful pedagogical review of different tools for studying aperiodic variability).

So, what are these underlying physical processes? Over the decades, much theoretical work has gone into understanding the variability processes at play in AGN. In recent years, radiation MHD simulations can even be run long enough to also participate in these studies, and promising work is beginning to reproduce the observed variability from first principles \citep{hogg16,turner2023, secunda2024}. To date, we still do not fully understand all the complex variability processes at play, but in short, the variability can generally be described by the theory of ``propagating fluctuations" \citep{Lyubarskii97}. Fluctuations in the disk's surface density are driven by the interaction between magnetic fields and differentially rotating gas (known as MRI; \citealt{Balbus1998}). These local fluctuations occur at all radii in the disk, and then evolve through the disk at the local viscous time (Eq.~\ref{eq:visc}). Fluctuations originating in the outer, cooler disk produce slower, lower-energy flares and then propagating inwards, later modulating the variability of faster, hotter independent flares from the innermost regions. The stochastic light curve we observe is then due to the response of the accretion flow and corona to these intrinsic fluctuations \citep{mummery2024,uttley2023}.

Here we review the basic properties of X-ray variability in AGN and compare to what is seen at longer wavelengths. We do not attempt to summarize all the rich phenomenology, but instead highlight some key observables that help us infer physical properties, like the black hole mass and accretion flow viscosity, geometry, and dynamics. 

\subsubsection{X-ray variability estimates the black hole mass} 

Even from very early X-ray observations of AGN light curves \citep{barr1986}, it was recognized that luminous AGN showed less variability than their lower luminosity counterparts. Now, we understand this observation as largely due to the fact that the characteristic variability scales with the black hole mass (Fig.~\ref{fig:times}). The dependence can be seen anecdotally in Fig.~\ref{fig:massPSD}-{\em left}, where we compare the {\xmm} $0.3-10$\,keV light curves of four nearby Seyfert 1s, spanning roughly four orders of magnitude in mass from $10^{5}$\,{\msun} to $10^{8}$\,{\msun}. In a 70~ks (20~hr) light curve, the largest black hole (Ark~120) shows variability on timescales of a few hours (corresponding to roughly the dynamical timescale for a $10^{8}$~\msun\ black hole, Eq.~\ref{eq:tdyn}), whereas for the same duration light curve, a $10^{5}$~\msun\ black hole (NGC~4395) shows variability on timescales as short as tens of seconds. In other words, for the same duration light curve, the smaller black hole probes variability on the thermal, viscous, and dynamical timescales. 

In a systematic study of Type 1 (unobscured) AGN in X-rays, \citet{Ponti2012} showed this behavior is common: the excess variance (a measurement of the intrinsic variance, where measurement noise is subtracted off) scales with the mass of the black hole on the timescales probed in a typical {\xmm} observation. As such, {\em measuring the amount of X-ray variability is a relatively inexpensive way to estimate AGN masses}. This has recently been confirmed in a larger number of sources across a larger luminosity range, and out to high redshifts \citep{elias-chavez2024,Prokhorenko2024}.

\subsubsection{The X-ray and UVOIR power spectral densities}

One of the most common tools for quantifying the average AGN variability properties is the power spectral density (PSD), which is a measure of the variability power (the mean of the squared amplitude) as a function of temporal frequency, 1/timescale \citep{green1993,lawrence1993}. This is a powerful tool for measuring how turbulence builds up in the accretion flow of standard AGN. Moreover, recently, large scale time domain surveys have been used to search for periodic or quasi-periodic signals from exotic transients (e.g., from binary supermassive black hole, or other orbiting objects, see Sections~\ref{sec:QPE} and \ref{sec:SMBHB}). In order to definitively find these exotic outliers, it is vital first to have robust understanding of the variability behavior of individual AGN across different masses and accretion states. Here, we review the X-ray and optical PSD properties of standard AGN.

Fig.~\ref{fig:massPSD}-{\em right} shows the corresponding X-ray PSDs for the four AGN light curves to the left. Seyfert 1s typically show high-frequency PSDs that are well described as a power-law, where the variability power $P(f)$ scales with frequency $f$ as $P(f) \propto f^{-\alpha}$. AGN light curves typically show so-called ``red noise'' power spectra, where the power-law slope of the PSD, $\alpha$, is $>1$. In a systematic study, \citet{Gonzalez-Martin2012} found that X-ray PSDs of Seyfert~1s in the frequency range of $10^{-5}$~Hz to $10^{-2}$~Hz (describing rapid variability from minutes to $\sim 1$~day) are generally consistent with $\alpha\sim 2$ (so-called ``random walk'' noise), albeit with large scatter across sources. Two light curves of the same AGN in a given accretion state  will look different because of statistical fluctuations (inherent to any stochastic process), but their power spectra will be the same if they arise from the same underlying physical process. 

The X-ray variability power does not keep rising indiscriminately at longer timescales, and instead at some frequency (the so-called break frequency), we observe a turn-over to a shallower slope. This break frequency scales with mass and accretion rate in AGN and even down to stellar-mass black holes \citep{Mchardy2006}. 
While the origin of the break frequency is unknown, it marks the timescale above which variability is produced less efficiently. This timescale has been associated with the viscous timescale at the inner edge of an optically-thick disk (e.g., a $10^{6}$\msun\ AGN may show a break frequency of $10^{-5}$~Hz or 27~hrs, which is roughly the viscous timescale at the  ISCO of a non-rotating black hole, Eq.~\ref{eq:visc}). The break frequency has alternatively been associated with the cooling timescale of electrons in the Comptonization model \citep{Ishibashi2012}.

\begin{figure*}
\begin{center}
\includegraphics[width=\textwidth]{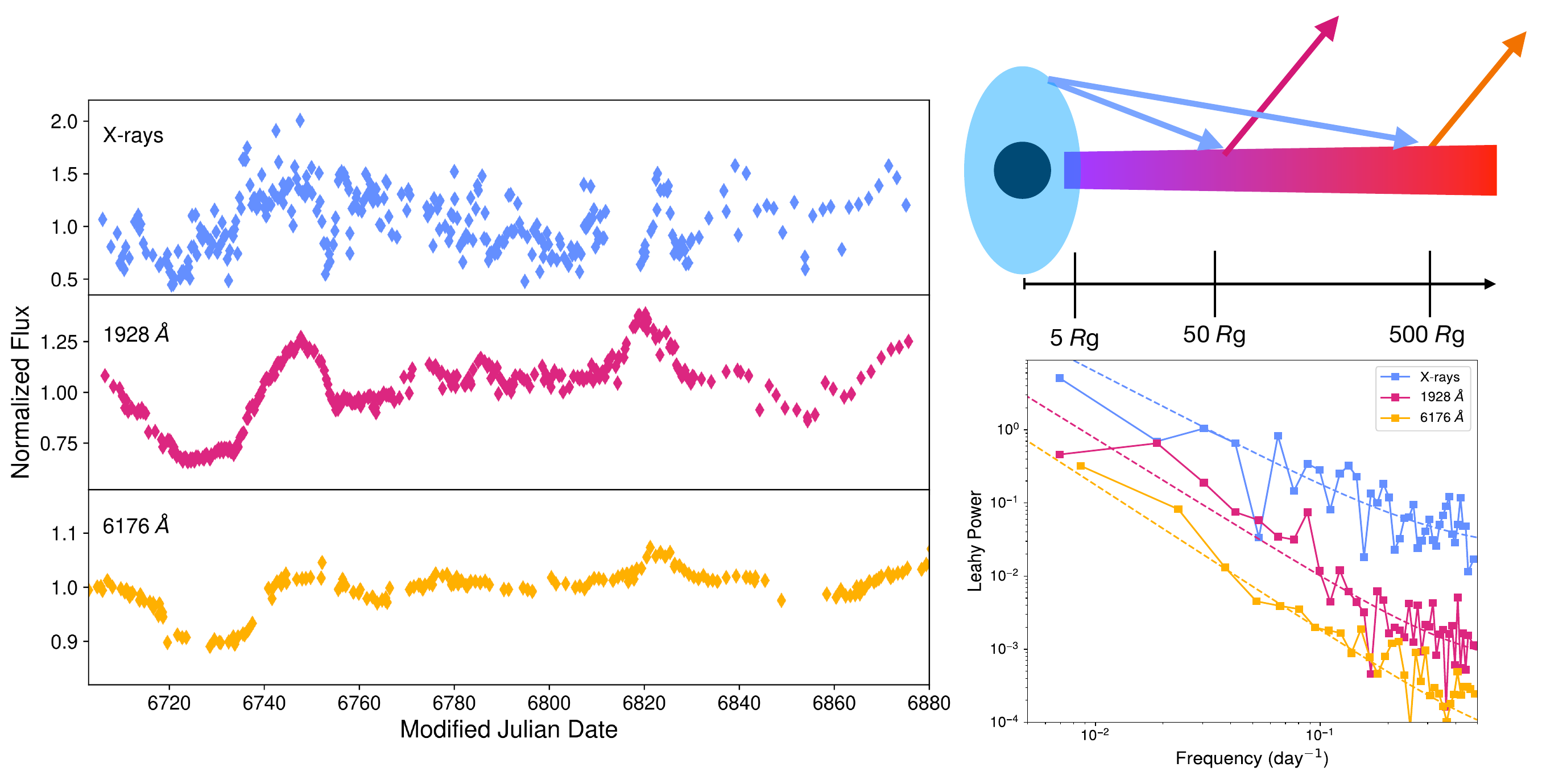}
\caption{{\it Left}: A comparison of the X-ray light curve (blue) to longer wavelength UV (pink) and optical (yellow) for the AGN NGC~5548. The X-rays are most rapidly variable, and originate at smallest scales (within 10~$R_g$ of the black hole). The UV and optical originate further out in the accretion disk. Most of the rapid variability observed in this $\sim 1$-year campaign is much too fast to be driven by changes in the mass accretion rate through the disk (Eq.~\ref{eq:visc}), but rather is due to irradiation of the disk by the highly variable X-ray corona. This irradiation from a central source explains why the UV and optical are delayed with respect to the X-rays (See Section~\ref{sec:reverberation}), and why the UV and optical power spectra (bottom-right) are steeper and lower normalization than the X-ray PSD (Section~\ref{sec:timing}). On the longest timescales, variability due to mass accretion rate fluctuations may also contribute in some cases (e.g., see \citealt{Hernandez2020,Hagan2024}). Light curves and PSDs for NGC~5548 courtesy of Christos Panagiotou (MIT).}
\label{fig:NGC5548}
\end{center}
\end{figure*}

Complimentary optical studies on sub-day timescales have been challenging, but the {\em Kepler} exoplanet-discovery mission did provide long uninterrupted light curves at high cadence (especially during its K2 mission phase), thus allowing for similar variability analysis on the same timescales as typically probed in the X-rays. 
The high-frequency optical PSDs from {\em Kepler}  exhibit lower normalizations and steeper power-law slopes than X-ray PSDs (e.g. $\alpha=[2.6-3.3]$; \citealt{Mushotzky2011,Edelson2014,Smith2018}), and are inconsistent with expectations from the random walk ($\alpha \sim 2$) seen at X-rays. Similar conclusions are drawn from longer timescale simultaneous X-ray, UV and optical monitoring campaigns from {\em Swift}, as shown for the AGN NGC~5548 in Fig.~\ref{fig:NGC5548} (adapted from \citealt{Panagiotou+2020}). While X-rays often vary by $>50\%$ day-to-day, the variability amplitude decreases to $\sim 20$\% in the UV, and $<10\%$ in the optical. Simply put: {\em optical/UV light curves do not show as rapid variability as X-rays, indicating that the X-ray corona is more compact than the optical/UV-emitting disk.}

In recent years, with the increasing time baseline and footprint of large all-sky optical time domain surveys, these studies can be expanded to large samples of AGN. While the sampling is poorer than in targeted campaigns, the number of AGN is enormous, and careful stacking shows that the optical PSD slope above the break is $2.5-3$ \citep{Arevalo2023}. This rejects the commonly used DRW model. 
\begin{marginnote}[]
    \entry{DRW}{{\it Damped Random Walk}, PSD with $P(f) \propto f^{-2}$ that breaks to flat slope at long timescales.}
\end{marginnote}
Similar to earlier X-ray studies, these authors find that the optical PSD amplitude decreases and slope steepens with increasing mass and Eddington ratio (\citealt{Arevalo2023,Petrecca2024}). This is perhaps related to the fact that the X-ray to bolometric luminosity ratio decreases with increasing Eddington ratio (see Section~\ref{sec:disk-corona}), and as such, higher luminosity AGN do not have enough X-ray emission to modulate the UV variability. Moreover, the disk temperature increases with mass accretion rate (Eq.~\ref{eq:fdisk}), and thus a fixed optical bandpass probes larger radii for higher accretion rates. Emission originating from a larger radius (i.e. larger emitting region) will naturally be less variable. This may be an important point in predicting the rest-frame UV variability of Little Red Dots, if indeed they are super-Eddington AGN (e.g. \citealt{pacucci24}).

Studies of the broadband noise properties of power spectra have largely been phenomenological, but there are promising physical models that attempt to self-consistently explain the variability across different energy bands \citep{papadakis2016}, e.g., by predicting how the X-ray variability from the corona is damped out as it irradiates the accretion disk producing reprocessed UV variability \citep{Panagiotou2022}. These types of physical models will become more constraining with the launch of missions like {\em ULTRASAT}, providing unprecedented high-cadence UV light curves on timescales of days to minutes for a large number of AGN.

\subsubsection{Low-frequency hard lags are ubiquitous}\label{sec:lflags}

\begin{figure*}
\begin{center}
\includegraphics[width=\textwidth]{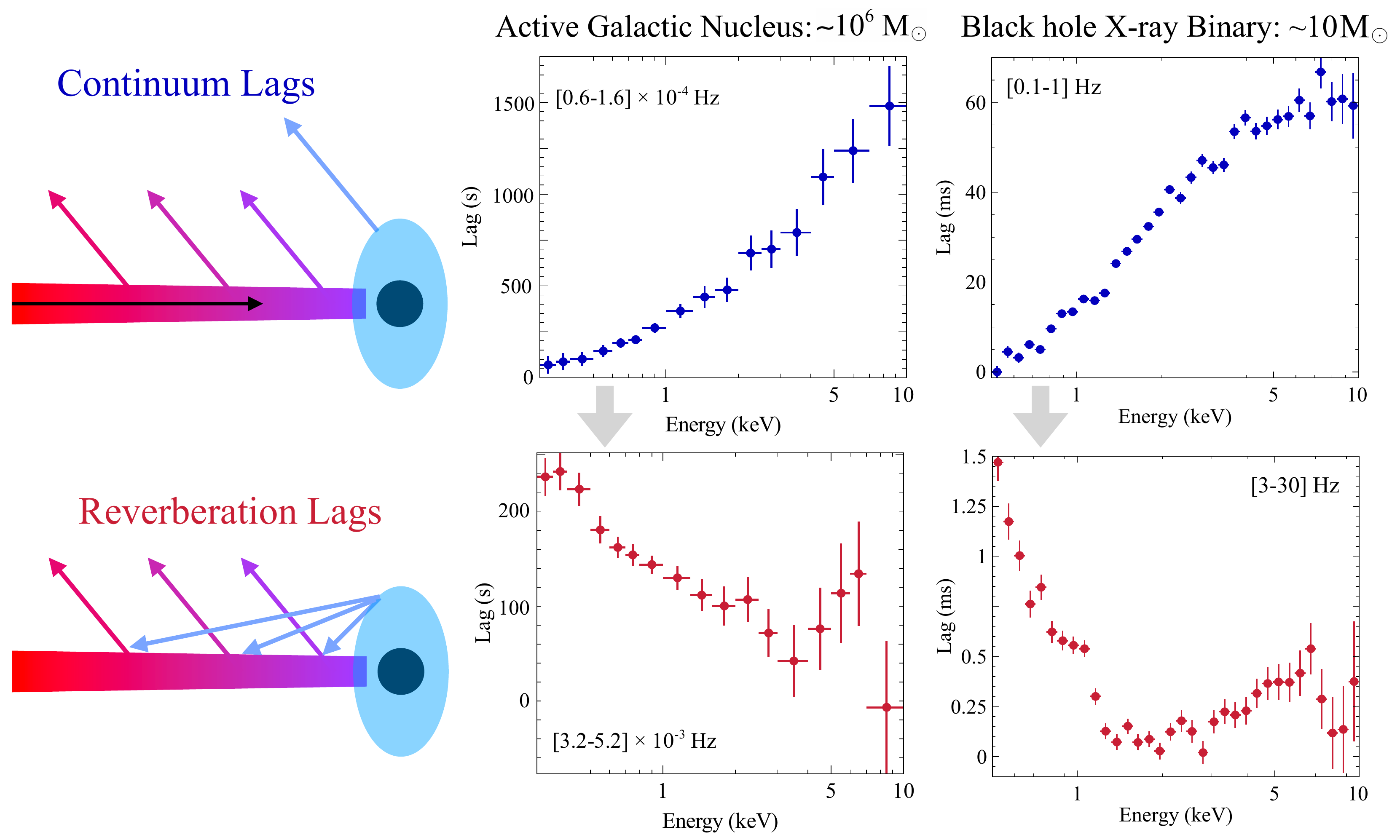}
\caption{A comparison of the frequency-resolved time lags for the $\sim 10^{6}$~\msun\ AGN Ark~564 (middle column), compared to the black hole X-ray binary MAXI~J1820+070 with a $\sim 10$~\msun\ black hole. The ``low-frequency" lags (top row) show a featureless log-linear increase with energy, and are understood to be due to mass accretion rate fluctuations propagating through the disk and into the corona. At high frequencies (bottom row), we observe a distinctly different dependence on the lags with energy. These lags are instead understood as due to reverberation, as the corona irradiates the cooler disc (Section~\ref{sec:reverberation} for details). The lags in luminous Seyferts look very similar to the lags in the luminous hard state of accreting stellar mass black holes, supporting the idea of mass invariance in accreting black holes (see Sidebar entitled {\it Stellar-mass Black Holes} for details). 
}
\label{fig:lags}
\end{center}
\end{figure*}

Long timescale hard-band lags are another fundamental property of accreting black holes, and encode information about the continuum source (Fig.~\ref{fig:lags}-{\em left}). In the context of AGN, `long' timescales are $\sim 5\times 10^{-5} c/R_g$ (or $10^{-5}$~Hz for a $10^{6}$\msun\ black hole). The observed hard-band lags were originally discovered in black hole X-ray binaries \citep{miyamoto1989, nowak99}, and later discovered in AGN \citep{papadakis01}. In \citet{kara16}, we showed that these hard lags were a ubiquitous feature in unobscured AGN systems. 

Because the variability is rapid, and a single uninterrupted observation with, for instance, {\xmm} can capture a range of variability timescales, time lags in X-rays are most commonly studied in the Fourier domain (see \citealt{nowak99,uttley14} for descriptions of the technique). Whereas the cross-correlation function (traditionally used in optical studies) provides a measurement of the time lag averaged over all variability timescales, the Fourier approach allows for the measurement of different timescales at different temporal scales. The benefit of this approach is that it allows one to measure lags from different physical mechanisms that dominate the variability at different timescales. For instance, in a standard unobsured AGN, when we zero in on the long-timescale variability (low-frequency), we observe the hard X-rays lagging behind the soft X-rays, but when isolating the short-timescale variability (high-frequency), we see the opposite--the soft X-rays lagging the hard. These high-frequency soft lags are interpreted as due to reverberation echoes, which will be discussed in Section~\ref{sec:reverberation}, but for now, we want to make the point that these two different physical lags can only be ``teased out'' of the data when using the frequency-resolved Fourier analysis, whereas a standard cross-correlation analysis averages over all variability timescales, and thus would only measure the low-frequency hard lags that dominate the variability power. 

When isolating variability on long timescales, we observe a featureless log-linear increase in time lag with energy, indicating that this is a property of the continuum emission. \citet{kotov01} first explained these lags within the context of the propagating fluctuations model: Mass accretion rate variations produced in the outer, {\em cooler} parts of the disk propagate inwards, leading to correlated, and delayed emission from fluctuations from the closer-in, {\em hotter} portions of the disk. The greater the difference in energy, the longer the lag, as the propagation front travels a larger distance through the disk. \citet{Arevalo2006} expanded these studies, and demonstrated that the same model developed for black hole X-ray binaries can apply to AGN, assuming that the corona tracks the variability of the thermal seed photons. It is unlikely that the observed lags are due to fluctuations that propagate through the corona itself, as the propagation timescale through the optically-thin corona is much shorter than the viscous timescale through the disk. Recently, \citet{uttley2023} demonstrated that fluctuations propagating through the disk increase the seed-photon flux entering the corona, thus cooling it, which leads to a steeper powerlaw shape, and as its re-heated (via e.g. magnetic reconnection) the spectrum hardens again. The result is a pivoting of the slope of the continuum spectrum as it fluctuates in luminosity, which leads to hard-band lags \citep[e.g.;][]{Ingram2019}.    

As of recent years, global MHD simulations of accretion disks can now be run long enough to test such propagating fluctuations models (\citealt{hogg16}, \citealt{turner2023}). They appear to rise from the magnetic dynamo. The amplitude and timescale of the variability is set by the microphysics of the accretion flow and the local viscous timescale. The simulations are beginning to reproduce observables like the PSD and the low-frequency hard lags \citep{turner2023}. In principle, by comparing observed variability properties to results of MHD simulations, we can probe properties of the accretion flow at the disk midplane that are otherwise unobservable. 


\subsection{X-ray Reflection Spectroscopy: A Probe of the inner 
accretion flow and black hole spin}\label{sec:reflection}

\begin{figure*}
\begin{center}
\includegraphics[width=\textwidth]{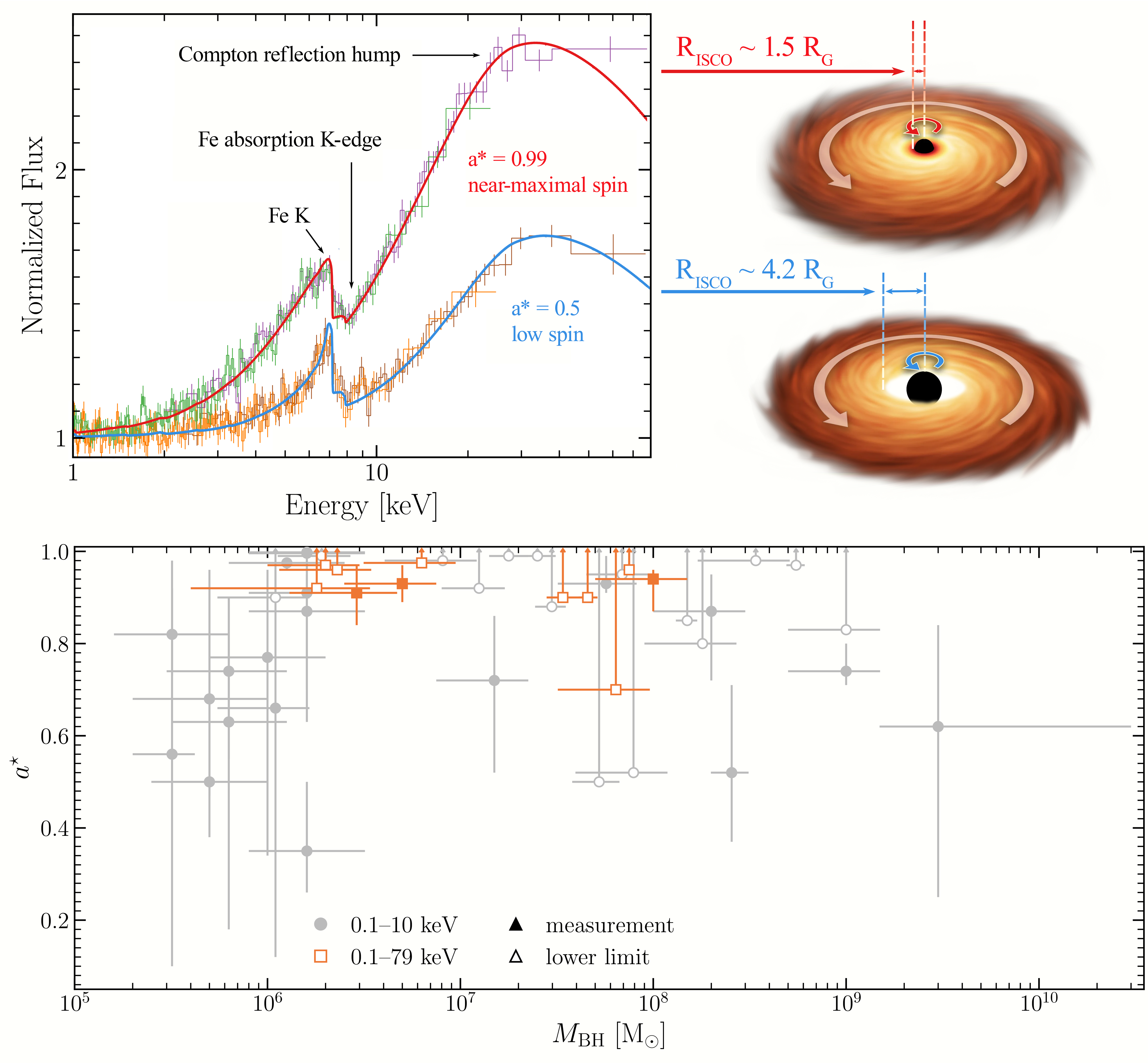}
\caption{{\it Top}: Demonstration of the effects of the black hole spin on the accretion disk ({\it right}) and thus in the reflected X-ray spectrum ({\it left}). Relativistic frame-dragging effects cause the inner-edge of the disk to move closer in (i.e., lower \risco) for rapidly spinning black holes, as shown in the artistic rendition on the {\it right}. The corresponding reflection spectra ({\it left}) from a {\xmm}+{\nustar} observation simulated with the \relxill\ model \citep{dau14,gar14a} show the stronger distortion of the spectral features when the emission is produced closer to the black hole (high-spin case). Figure adapted from \citet{Garcia2024}.
{\it Bottom}: Distribution of SMBH spins as a function of their mass constrained observationally using X-ray reflection spectroscopy. Filled and open symbols differentiate full constraints from lower limits. Orange and gray data points indicate measurements using a broadband ($\sim 0.1-78$\,keV) or soft-energy data only ($\sim 0.1-10$\,keV), respectively. Adapted from \cite{Piotrowska+2024}.
}
\label{fig:spins}
\end{center}
\end{figure*}

For relatively luminous systems ($L_{bol} \sim 0.01-0.3 L_{\mathrm{Edd}}$)  a fraction of the emitted photons irradiate the surface of the disk, producing a {\it reflected} emission component that contains a diverse range of atomic spectral features. Among these, the inner-shell transitions from Fe ions---predominantly the K$\alpha$ emission lines at $6.4-6.97$\,keV (depending on the ionization state)---together with a strong absorption K-edge around $7-8$\,keV, are typically the most prominent. Furthermore, reflection spectra exhibit a characteristic high-energy continuum, peaking at $\sim$30\,keV, known as the {\it Compton hump} \citep{George91, ross05, Garcia+2020}. Although the emission lines are narrow in the frame of the disk, once the accreting material approaches the ISCO, it is orbiting at relativistic speeds, while the emitted radiation experiences a strong gravitational redshift. These spectral components are shown in the top-left panel of Figure~\ref{fig:spins}. Together, these effects both broaden and skew the emission lines from our perspective as a distant, external observer, resulting in a characteristic ``diskline" emission profile \citep{Fabian1989, kdblur, kerrconv, relconv}.
\begin{marginnote}
    \entry{Diskline}{
    Broad and asymmetric line profile emitted by a disk that is smeared by Newtonian and relativistic effects.
    }
\end{marginnote}

Provided the disk extends down to the ISCO, as expected at moderately high accretion rates, then characterizing the relativistic blurring gives a measurement of $R_{\rm{ISCO}}$, and consequently provides an estimate for the black hole spin (Fig.~\ref{fig:spins}, top-right). This can be done by virtue of the relationship between \risco\ and the dimensionless spin parameter \spin$=Jc/GM^2_\mathrm{BH}$ \citep{Bardeen1970}. The location of \risco\ changes by almost a factor of 10 between the two possible extreme spin values \citep[i.e.; \spin$=\pm 0.998$, for maximal prograde or retrograde spin, respectively;][]{Bardeen+1972,Thorne1974}.

There are a variety of methods to recover the black hole spin, but the so-called {\it relativistic reflection} model is the most successful method that can be reliably applied to AGN. While relativistic effects impact the entire reflection spectrum (Fig.~\ref{fig:spins}, top-left),  historically, this method is also referred to as the {\it Fe-line method}, as the K-shell iron emission is typically the strongest spectral feature. 

In fact, broadband X-ray spectroscopy (beyond just fitting the iron line) is imperative for robust spin parameter constraints because the broadband continuum modeling (from $\sim 0.3-50$~keV) must be accurate \citep[see][for  a recent review]{Reynolds2021}. Now, since the launch of the hard X-ray telescope \nustar\, and in coordination with soft X-ray observatories, we have those broadband high signal-to-noise spectra, which has allowed for determination of black hole spin in a large fraction of AGN with such high S/N observations \citep{Risaliti13nat, Ricci14, Wilkins15, Jiang18iras, Walton19ufo}.

Figure~\ref{fig:spins} (bottom) shows the current state-of-the-art of X-ray reflection spin measurements for AGN as a function of black hole mass, based on the most recent compilation of \cite{Piotrowska+2024}. This list of sources has been updated from the initial compilation of \cite{Reynolds2021}, which placed several quality criteria to select sources from all reported values in the literature. The current list contains 46 sources ranging over five decades in mass. For half of the sample, only lower limits can be imposed in their spin, given the limited signal of the data. Moreover, only 11 sources have spins measured over a broad-band spectrum (i.e., including $3-78$\,keV data  from \nustar, where most of the reflection features are observed). Thus, it is possible that some of these measurements are biased, as they might depend strongly on the model fit to the soft energies only (i.e., fitting the soft excess with reflection; see Sect.~\ref{sec:soft_excess}).

The spin uncertainties shown in Figure~\ref{fig:spins} are only statistical, and do not include any possible systematic uncertainties inherent to the modeling of the reflection spectrum, which might include: accretion disk geometry, e.g., most models assume a razor-thin disk, while a vertical extension can influence the shape of the reflection signatures \citep{taylor18}; physics of the accretion flow, which implies understanding the density profile in both vertical and radial directions (e.g., constant density vs. hydrostatic solutions vs. full radiation MHD simulations); and coronal geometry, which affects the illumination pattern produced by a simplified lamppost geometry vs. more realistic ones \citep[e.g.,][]{Niedzwiecki+2008,Wilkins2012a}. See reviews by \cite{Reynolds2021} and \cite{Bambi+2021} for additional discussions. Nevertheless, accurate spins can be determined by analyzing multi-epoch and broadband observations of the same source, combined with careful comparisons of different model flavors (i.e., different geometries).

In the last few years, reflection models have become increasingly more physically motivated, e.g., by producing fully self-consistent models assuming a lamppost geometry \citep[e.g.;][]{dauser13}, significantly improved microphysics by including high-density plasma effects in the atomic data and ionization calculations \citep{gar16b,Kallman+2021,Ding+2024}, and other relativistic effects such as disk self-illumination \citep{Connors2020}, multiple reflections \citep{Wilkins+2020}, returning radiation \citep{Dauser2022}, and both thermal \citep{Mummery+2024} and reflected \citep{Dong+2023} emission from the plunging region.

The relativistic reflection approach to measuring spin is particularly powerful because it can be applied to black holes of all masses, not just in AGN \citep[e.g.;][]{Walton12xrbAGN}. Measuring spins in black hole X-ray binaries offers a way to test the systematic uncertainties of reflection modeling because in these lower mass systems, we have other methods for estimating black hole spin, the most well-studied of which is the so-called {\it Continuum Fitting Method}. In X-ray binaries, the accretion disks are so hot that they emit X-rays. Moreover, during an outburst, they conveniently transition to the ``soft state", where the corona is weak, the disk dominates and we have a clean view of the Wein tail. By accurately measuring the thermal spectrum, we can estimate the inner radius of the disk, which, if associated with the ISCO, determines the spin \citep{Zhang+2019,McClintock+2014}. From the five systems for which both methods have been applied, there is agreement in all except one, GRO~1655$-$40, for which the constraint from reflection ($>0.9$) is larger than that from the continuum fitting \citep[$0.7\pm0.1$, see Table~2 in][]{Reynolds2021}. 

Measuring the spin of stellar mass and supermassive black holes ultimately informs our understanding of the growth history of these black holes, and how much accretion via the disk can add angular momentum to the black hole over time (e.g., see \citealt{volonteri05} for AGN, and \citealt{fragos15} for low-mass XRBs). The current sample of SMBHs is not yet sufficient to fully (and statically) discriminate among different evolutionary models for SMBH growth, nor does it provide enough constraints to directly inform cosmological simulations \citep{Piotrowska+2024}, but the prospects are encouraging. {\it The community faces the challenge of coordinating large observational campaigns in tandem with the most sophisticated reflection models to increase both the sample size and its accuracy.}

\subsection{Reverberation Mapping: A probe of the disk and corona geometry}

\subsubsection{X-ray Reverberation and the confirmation of the relativistic reflection model}\label{sec:reverberation}

The discovery of X-ray reverberation lags in AGN was a critical confirmation of the relativistic reflection model \citep{fabian09,zoghbi10}. One major prediction of the relativistic reflection scenario presented above is that the reflection spectrum should be delayed with respect to the continuum due to the additional light-travel path between corona and disk. This is similar to the general idea of optical broad line region reverberation lags, discovered in the 1980s, and which is commonly viewed today as our best means for measuring the masses of SMBHs (e.g., \citealt{peterson04}). In optical broad line reverberation mapping, one cross-correlates a light curve dominated by the primary continuum (in this case disk emission from hundreds of gravitational radii) and a light curve of the broad Balmer lines (produced via fluorescence of material further out in the BLR). Similarly, in X-rays, we search for the time delays between light curves dominated by the primary coronal continuum and those dominated by the reflection spectrum. As the emission lines in the X-ray spectrum are much broader than in the optical, this does add the additional complication that no band is {\em purely} continuum nor {\em purely} reflection. As such, careful modeling of the lags with general relativistic ray-tracing simulations is required to infer the geometry of the inner accretion flow (e.g., \citealt{cackett2014,chainakun2016,Ingram2019}). Moreover, in the X-ray band, the dominant time lag occurring on long timescales is the hard band lag (associated with the continuum and interpreted as due to propagating mass-accretion rate fluctuations; see Section~\ref{sec:lflags}). One only directly observes the reverberation time delays by doing a Fourier de-composition to remove the low-frequency hard lags, and isolating the reverberation lags at high-frequencies (\citealt{uttley14} and see Fig.~\ref{fig:lags}). 

While the timing technique and modeling is somewhat more complicated than required in traditional optical BLR measurements, this technique has been successful in measuring X-ray reverberation lags in about two dozen sources that are bright enough, have sufficient rapid variability, and are not highly obscured \citep{demarco13,kara16}. For comparison, for a typical $10^{7}$\msun\ black hole, one finds that the BLR light curve lags the continuum typically by $\sim 1$~week  \citep{cackett21}, corresponding to light-travel distances of centi-parsecs ($\sim 10^{5}~R_g$). In X-ray reverberation, for a similar mass black hole, one measures time delays between the continuum and reflected light curves that are $\sim 100$~seconds, corresponding to light-travel distances on the order of $\sim 10~R_g$. 

The fact that one can self-consistently model both the time-averaged spectra and the time-dependence of those spectra was an important step in ruling out early competing models that suggested that the inner edge of the accretion disk was highly truncated and thus could not be used as an indicator of black hole spin (e.g. \citealt{mizumoto14}). In these alternative models, circumnuclear gas at hundreds/thousands of $R_g$ from the black hole (perhaps from the inner BLR) caused absorption along our line of sight that could mimic the shape of a relativistically broadened iron line. However, such a model cannot explain the short lags associated with the soft excess and iron~K line. Moreover, the same short lags are seen in black hole X-ray binaries, where there is a disk, but no a BLR \citep{demarco17,Kara19}. {\em The short X-ray reverberation lags confirm that the disk is not truncated, and that the X-ray corona is a distinct plasma confined above the surface of the disk.} The reverberation results (e.g. \citealt{zoghbi12,kara16}) and the steep emissivity profile inferred from reflection spectroscopy (e.g. \citealt{Jiang19agn}) both suggest that the corona is radially confined to within $\sim 10 R_g$, though recent \ixpe\ observations appear to suggest somewhat greater radial extent (Sec.~\ref{sec:polarization}). {\em Future modeling efforts that simultaneously model spectral, timing, and polarization data are required.}

\subsubsection{Connecting the X-rays to longer wavelengths}

Because X-ray reverberation requires different telescopes, techniques, and models than traditional optical BLR reverberation mapping, these two fields developed in parallel for several years, but recently, there have been major efforts, like those of the AGN STORM 2 collaboration \citep{Kara2021}, to bring these communities together, to reverberation map the gas flows around AGN from the torus (as probed via IR; Landt et al. in prep.), down to the broad line region (with UV and optical spectroscopic; \citealt{homayouni23}), to the outer accretion disk (as probed via optical and UV continuum; \citealt{cackett23}), and all the way down to the ISCO (via X-rays; \citealt{lewin24}). For more details on each of these different reverberation techniques, and what they tell us about different scales of accretion, see the recent review by \citet{cackett21}. Reverberation mapping from the torus to the ISCO is an enormous community effort, and the campaigns are very challenging to get all the data with as few observing gaps as possible, but when successful, they provide an unprecedented look at the lifecycle of an AGN: how gas flows funnel from the surrounding galaxy, through the disk and towards the black hole, and in doing so, launches outflows that may constitute the BLR itself, and may be powerful enough to affect the galaxies at large scales (see Section~\ref{sec:outflows}).

Multi-wavelength reverberation mapping campaigns from X-ray to NIR have been performed on roughly ten AGN, and while each AGN has its own idiosyncrasies, a few general results have been found: In all cases, the optical lags behind the UV, following the functional form that the time lag $\tau \propto \lambda ^{4/3}$, as expected for irradiation of a disk around a centrally-located ionizing source (e.g. \citealt{edelson19}). This, together with the fact that the X-rays are more rapidly variable than the UV (see Section~\ref{sec:timing}.2 on the multi-wavelength PSDs) is the best evidence that the  weeks to months timescale variability observed in the UV/optical originates from reprocessing of X-rays from corona. The disk does produce its own variability due to turbulence (see Section~\ref{sec:timing}.4 on propagating fluctuations), but at the radii where the optical and UV peak, this variability is much slower (hundreds of years), and is not typically observed (though see \citealt{Hernandez2020} and \citealt{yao2023} for detection of propagating fluctuation lags in Fairall~9, seen only by removing more rapid variability associated with reverberation).  

While the corona-disk reverberation picture generally holds, there are caveats: We uniformly observe that the amplitude of these lags are several times longer than expected for a standard Shakura-Sunyaev thin disk \citep{edelson19}. This, together with the fact that the U band (containing the Balmer jump from the BLR) shows an excess lag inconsistent with just reverberation from the disk, has led to the conclusion that there is some contribution from the BLR in these disk lags (although the exact contribution of the BLR lag is still debated; \citealt{Kammoun2019,chelouche19}). One solution to the ``too-long-lag" problem is to have a better treatment of the ionizing source. If one accounts for the geometry and energetics of the variable corona, and the fact that some of the light that irradiates the disk is reprocessed as fluorescence lines (what we often refer to as ``X-ray reflection"), while some is re-emitted in the optical/UV, it can nicely explain the observed lag amplitude \citep{Kammoun2019}, PSD shapes and normalizations \citep{Panagiotou2022}, SED \citep{Kammoun2024}, and the amount of correlation between X-ray and UVOIR light curves \citep{Panagiotou2022-correlation}. X-ray reverberation lags also provide independent constraints on the geometry and energetics of the corona, and so {\em future work modeling both X-ray and UVOIR disk lags will be an important step.}

\subsection{The ubiquitous soft excess, and its debated origin}\label{sec:soft_excess}

\begin{figure*}[t]
\begin{center}
\includegraphics[width=\textwidth]{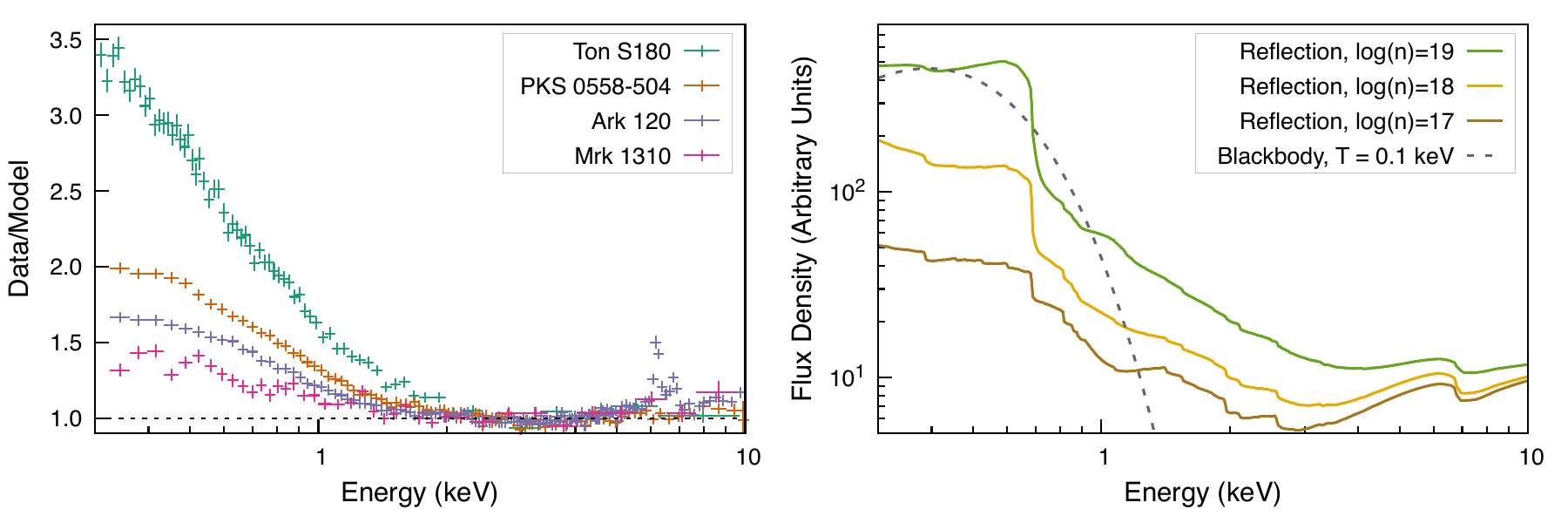}
\caption{Left: {\xmm} observations of four representative AGN displaying different amounts of soft-excess emission at low energies, shown as a ratio of the data to a power-law continuum model. Right: Reflection spectra calculated with the \texttt{relxillCp} model for different gas densities, compared to a canonical $T=0.1$\,keV blackbody component typically used as a phenomenological description of the soft excess. \xmm\ data courtesy of Elias Kammoun (Caltech).}
\label{fig:softE}
\end{center}
\end{figure*}

In a large fraction of Seyfert 1 AGN \citep[at least $\sim 50$\%, e.g.;][]{Boissay+2016} a soft excess component is observed peaking below $\sim 1-2$\,keV (Figure~\ref{fig:softE}, left). Its origin has been discussed over the years. Initially this soft excess was believed to be the hard tail of UV blackbody emission from the accretion disk \citep{sin85,arn85,pou86,mag98,lei99}; however, this explanation was ruled out given that systems with very different accretion rates and/or masses could be characterized by the same blackbody temperature, which is not expected for an accretion disk \citep{gie04,por04,pic05,min09,Bianchi+2009}.

The current models that have gained the most traction in the last years are: (1) Comptonization of UV photons in a ``warm" corona, and (2) ionized-blurred reflection. In the first case, the disk photons are Comptonized by an extended warm corona above the disk, which is optically thicker ($\tau\sim10-40$) and cooler ($T\sim10^6$\,K) than the corona responsible for the primary X-ray emission \citep{cze87,mid09,jin09,don12}. In the second case, the emission lines produced by the disk are relativistically blurred due to the proximity to the black hole \citep[see Sec.~\ref{sec:reflection}. Also see][]{cru06,fabian09,gar10,wal13,vas14}. 

\subsubsection{The Warm Corona Model for the soft excess.}
For many sources, the performance of these two models in describing the observed data is very similar, although often the warm corona leads to modest improvement in fit statistics \citep[e.g.;][]{Waddell+2023}. This is not surprising, given that in this picture the soft-excess is modeled with a thermal Comptonization model, which is independent of any of the other model components, particularly of the X-ray coronal emission. Thus, the model has complete freedom to adjust the spectral components independently, acting mostly as a phenomenological description. The thermal Comptonization models often used do not include any source of opacity other than electron scattering by design. On the other hand, models produced with full collisional excitation or photoionization equilibrium predict very strong absorption features in the spectra, which are not observed \citep{Garcia19}. Recent calculations performed with the TITAN/NOAR code by \cite{Petrucci+2020} including an additional source of (mechanical) heating predict featureless spectra. This additional heating presumably raises the gas ionization and diminishes atomic opacities. While this model has been successfully applied to a sample of AGN \citep{Palit+2024}, it is unclear how a plasma sufficiently ionized such that no atomic features are imprinted in the spectra, can at the same time produce a continuum with a relatively low temperature. More importantly, the physical origin of this mechanical heating is not specified. 

\subsubsection{The Relativistic Reflection Model for the soft excess}
On the other hand, the reprocessed radiation emergent from the disk---the reflection spectrum---also predicts excess soft-energy flux in the form of a forest of fluorescent emission lines from low-charge elements, or in the form of a bremsstrahlung continuum produced by free-free emission. The latter is significantly enhanced when the gas density in the disk is high \citep[$\gtrsim 10^{18}$\,cm$^{-3}$;][]{garcia16}, which are expected densities for accretion disks around black holes \citep{Shakura1973}. In fact, high-density reflection models have been used to describe the soft excess in several AGN \citep[e.g.;][]{jia18, mal18, Garcia19, Jiang+2019}, and also provide a self consistent explanation for the short soft lags discussed in Sec.~\ref{sec:reverberation}). Despite these successes, in several cases the model parameters are pushed to extremes, such as very high densities (so the soft X-ray flux is enhanced), and/or very high spins (so the spectral features are relativistically blurred). While these extreme parameters do not strictly violate any physical arguments, it could indicate a possible bias in the spin measurements of sources with strong soft excess, particularly in cases where only soft-energy data is included in the analysis (Sec.~\ref{sec:reflection}). The newest generation of reflection models include high-density plasma effects in the atomic parameters that had been previously neglected \citep{Ding+2024}, and show significant departures from standard models in the high-density regime. Further testing of these models on observational data hope to validate the role of reprocessed emission in explaining the soft excess. 

\subsubsection{Observational Challenges}
In general, any model that aims to explain the soft excess must address the surprising lack of correlation between its peak energy and the black hole mass or accretion rate, clearly demonstrated observationally by \cite{Bianchi+2009} (see their Fig.~9), while also predicting different possible strengths for the flux in the soft band. Figure~\ref{fig:softE} shows a comparison of four AGN with different strengths of the soft excess. Given the limited soft-energy band, it is not possible to distinguish if the strength of the soft excess is lower, or if its peak energy is out of the bandpass. Reflection models calculated at high densities produce enough flux below $\sim1$\,keV capable of mimicking this trend, which would explain the lack of correlation of the peak emission with the black hole parameters. However, the disk density is in fact a function of both mass and accretion rate, and thus there should exist a correlation between black hole mass and/or accretion rate and the strength of the soft excess. Future studies should test the presence of such a correlation.

The assumptions made about the origin of the soft excess directly affect our interpretation of the broadband X-ray spectra from AGN, particularly if related or not to the reprocessed disk emission. The answer to this question can have serious implications on the black hole spin estimates for a large fraction of sources, and consequently in constraining the spin distribution of SMBHs in the local universe (Sec.~\ref{sec:reflection}). {\it Thus, understanding the origin of the soft excess remains as a relevant issue in AGN physics to be addressed.}

\subsection{Narrow Emission Iron lines: A Probe of the BLR and Dusty Torus}\label{sec:blr-torus}

Any cold, dense gas around the black hole will fluoresce as it is irradiated by the X-ray corona. So far, we have focused on the corona irradiation of the accretion disk, where fluorescence lines produced at $\sim 10~R_g$ are broadened to several keV. But beyond fluorescence from the inner disk, there are also clear signatures of narrow ($<100$~eV; $<10,000$~km/s FWHM) fluorescence lines from larger scales \citep{Mushotzky1978} that are seemingly ubiquitous in AGN \citep{nandra2006}. These fluorescence lines were originally unresolved in X-ray spectra, making it difficult to estimate the distance to the emitting region, but the dusty torus at $\sim 1$~pc was the presumed emitter, as it was posited to cover $\sim 50$\% of the central source (leading to the 50--50 dichotomy of Type~1 and Type~2 AGN; \citealt{Antonucci1993}; see the review by \citealt{Hickox+Alexander2018}). 

The launch of {\chandra}/HETG \citep{canizares2000}, with $\sim 30$~eV energy resolution in the Fe~K band (several times better than typical CCD spectra), revealed that some of these narrow Fe~K lines in fact had line widths of several $1000$~km/s \citep{yaqoob2001,constantini2007, miller2018}, more suggestive of velocities associated with the optical BLR or outer disk. These spectroscopic measurements are corroborated by narrow Fe~K reverberation measurements, where one measures a time lag between the X-ray continuum and the narrow Fe~K fluorescence line that is on the order of several days \citep{ponti2013,zoghbi2019}, consistent with the reverberation time delays in optical H$\beta$ reverberation lags \citep{noda2023}. In general, it is understood that at least some of the emission line flux originates from size scales that are well within the dust-sublimation radius of the torus (\citealt{shu2010,gandhi2015} for earlier works), though the exact geometry, inclination-dependence and how these complex structures change between different systems is still not well understood \citep{Andonie2022}.  

\subsubsection{Narrow lines in the XRISM Era}

\begin{figure*}
\begin{center}
\includegraphics[width=\textwidth]{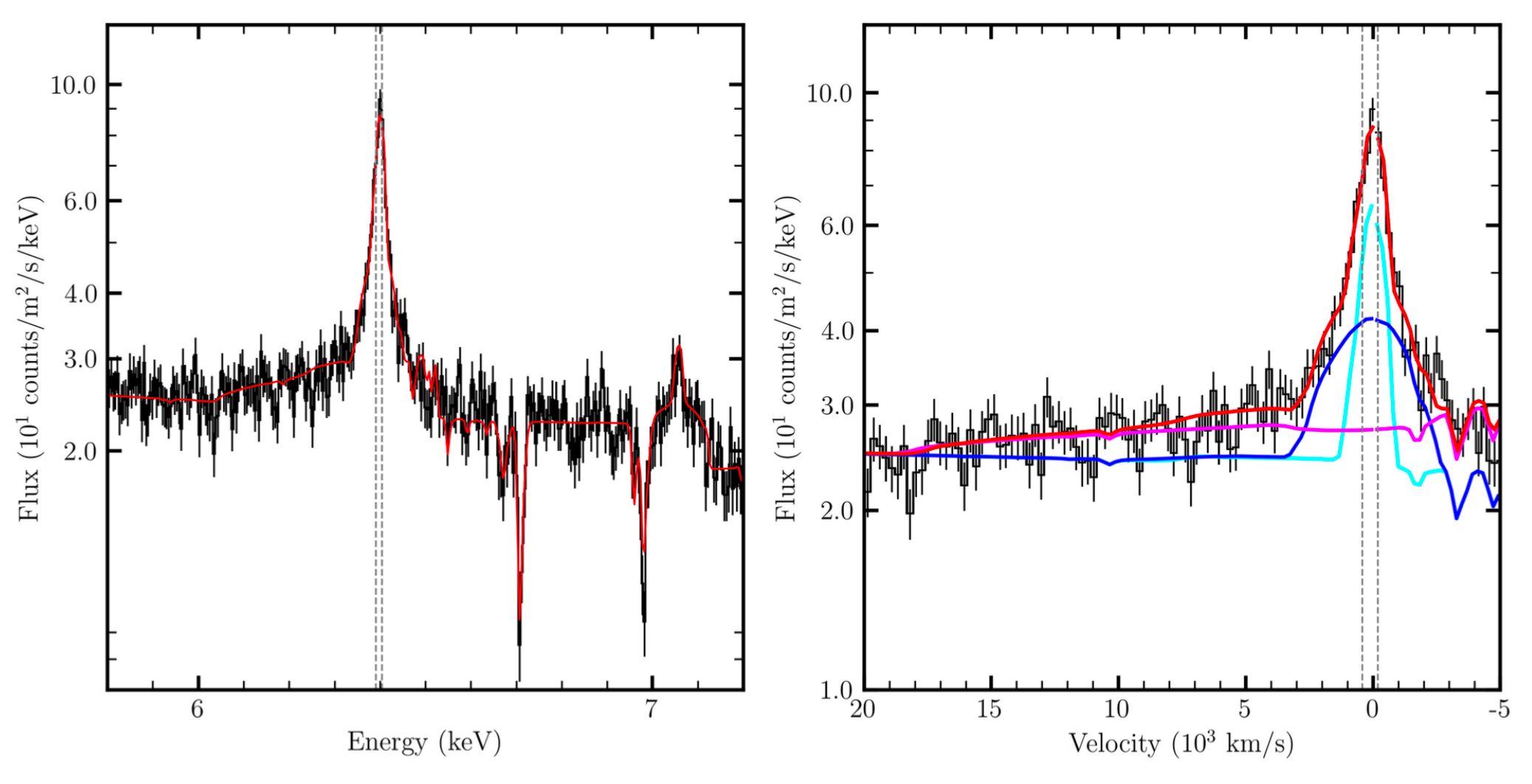}
\caption{{\em Left: }The {\xrism} spectrum of NGC~4151, zoomed in to the iron line region. The narrow emission line is resolved, and consists of three components, with velocities {\em right} consistent with the Keplerian velocity at the torus (cyan), inner broad line region (blue) and outer accretion disk (red). See Section~\ref{sec:blr-torus} for details. In addition to the emission line, there are absorption lines from Fe~{\sc xxv} and Fe~{\sc xxvi} transitions, blueshifted by a few hundred km/s (Section~\ref{sec:outflows} for details). Figure from \citet{xrism-4151}.}
\label{fig:xrism}
\end{center}
\end{figure*}

Now with the 2023 launch of {\xrism} \citep{tashiro2018}, observations of resolved Fe~K lines, which had been challenging and resource intensive for {\chandra}, are now easy and straightforward. {\xrism}/Resolve has an energy resolution of $\Delta E=4.5$~eV across the $1.6-17.4$~keV bandpass, providing a resolving power of $R=E/\Delta E = 1420$ or $\Delta v=200$~km/s FWHM at the Fe~K line ($\sim 6.4$\,keV). Indeed, the first results for an AGN published by the {\xrism} Collaboration shows that NGC~4151 (the brightest AGN in the sky) has a very complex iron line profile (\citealt{xrism-4151} and see Fig.~\ref{fig:xrism}). Assuming it fully originates from the blend of Fe~K$\alpha_{1}$ and Fe~K$\alpha_{2}$ at restframe energies of 6.404~keV and 6.391~keV, it requires broadening at three different velocities that can be associated to distances assuming Keplerian motion. This suggests emission from the torus ($\sim 10^{4}~R_g$), the BLR ($\sim 10^3~R_g$), and the accretion disk ($\sim 10^2~R_g$). This analysis was limited to the $6-6.6$\,keV range, and thus was not sensitive to even broader lines that could originate from the inner disk. What is key about {\xrism}, is that beyond estimating line widths via simple Gaussian modeling, we can now definitively measure the profile of the emission lines (e.g.; \citealt{miller2018}), including (relativistic) Doppler shifts, which will allow for high-fidelity estimates of the inclination and geometry of the emitting region. In addition to more sophisticated modeling of the dynamics of the gas, we similarly require a more detailed treatment of the atomic physics \citep[e.g.;][]{Kallman+2021,Mendoza+2021}, including proper treatment of the ionization state of the gas \citep[e.g.; with codes such as {\xillver;}][]{gar13a}. {\em {\xrism} measurements (especially together with optical spectroscopic and reverberation measurements, and interferometric observations with the GRAVITY VLTI that spatially resolve the BLR) offer a promising future to self-consistently map gas flows from parsec scales down to the inner accretion disk.}

\subsection{AGN Outflows: The search for AGN Feedback in action}\label{sec:outflows}

\subsubsection{X-ray Absorber: From slow outflows to ultrafast disk winds}

The X-ray spectrum also imparts information about optically-thin gas around the nucleus. The material is assumed to be distributed equatorially, and when intercepts our line of sight, we see its signatures in absorption. Typically this material detected via absorption is ionized (either photoionized from the central source or collisionally ionized), and often is observed to be outflowing at hundreds of km/s, and sometimes up to $0.1-0.4\,c$. 

AGN outflows are essential in understanding the role of SMBHs in galaxy evolution, as they can induce or suppress star formation, expel gas from the host galaxy, or halt accretion \citep{silk1998, Fabian2012}. Theoretical models infer that outflows with kinetic luminosities ranging from $0.5–5$\% Eddington luminosity of the black hole are necessary to have a substantial impact on galaxy evolution \citep{dimatteo2005,hopkins2010}, and thus there is much interest to understand the energetics and launching mechanisms of these outflows. However, to date, observational evidence of outflows capable of generating such feedback has been elusive.

AGN outflows manifest in several ways and on many scales: from the narrow-line region (NLR) and outer BLR, known in the literature as ``warm absorbers" with low velocities ($<1000$~km/s; \citealt{reynolds1995,behar2001}); to outflows at the inner BLR/outer accretion disk, sometimes referred to as ``obscurers" ($1000-5000$~km/s; e.g.; \citealt{mehdipour16}); to fast outflows launched at the inner accretion disk, called ``Ultrafast Outflows" (UFOs; e.g.; \citealt{pounds03}) that can reach velocities of  $0.1-0.4\,c$.
\begin{marginnote}[]
    \entry{Warm Absorber}{Slow outflows of $<1000$~km/s that typically originate at $\sim$parsec scales.}
\end{marginnote}
The literature is vast, and the nomenclature for different ``types'' of outflows can be jarring for the uninitiated, but it is possible that the outflows that we observe are all part of the same streamline, where UFOs launched from the inner regions eventually manifest as the slower, less ionized warm absorbers at larger scales \citep{Tombesi2013}. 
\begin{marginnote}[]
    \entry{Ultrafast Outflows}{Powerful, relativistic outflows of $>0.1$c launched from the inner accretion disk.}
\end{marginnote}
It is even posited that the optical BLR is a ``failed'' accretion disk wind, where the outflow velocity did not exceed the escape velocity, and falls back towards the disk \citep{czerny2011}. Understanding where these outflows are launched, how powerful they are, and if/how they couple to outflows at large scales (even out to galaxy-scale flows seen in sub-mm, IR and optical) is the subject of much important ongoing work. We invite interested readers to see the recent reviews by \citealt{gandhi2022} and \citealt{gallo2023} for more details.

\subsubsection{Future prospects for AGN Outflows}

Ultimately, we want to understand how the growth of the central supermassive black hole via accretion leads to large scale coupling to the growth of the galaxy, and to answer this question, not only do we need to understand which of these outflows are energetic enough to be responsible for large-scale feedback, but also how the accretion of matter on to the black hole launches these outflows. This is where spectroscopic time-domain campaigns are particularly powerful. If we can monitor the stochastic nature of the ionizing source, we can use the evolution of the spectra and the reverberation lags to understand the accretion disk and corona geometry, and subsequently, we can follow how the outflows change in response to these variations in the central engine. We can track how long it takes for the ionization state of the gas to respond to changes in the ionizing luminosity (as the recombination timescale of the outflow is very sensitive to the outflow density, a major uncertainty in determining the launching radius; \citealt{Rogantini+2022,Sadaula+2023}). By tracking the outflow evolution in time, we can test if the outflow velocity is set by the amount of continuum flux (as expected from radiation driven winds), or if more complex physics (i.e. magnetic fields) are at play. There have been attempts at coupling long-term multi-wavelength reverberation mapping campaigns with spectroscopic studies of the outflows (e.g. \citealt {DeRosa2015,Kara2021}). With improved time domain surveys (like {\it Rubin} in the optical and {\it ULTRASAT} in the UV), coupled with spectroscopic monitoring in the UV (with e.g. {\it UVEX}) and the X-ray (with, e.g. {\xrism}), we can probe the causal connection between accretion and ejection to answer some of the fundamental questions about the role of black holes in galaxy evolution.

\section{EXTREME VARIABILITY IN SUPERMASSIVE BLACK HOLES}\label{chap:variability}

As reviewed in the last section, AGN are extremely variable in X-rays. Because of their stochastic nature, we expect that each time we point our telescopes, we will see a different flux and a different variability pattern. But sometimes, this variability is beyond random fluctuations drawn from the same underlying physical process, and instead indicates a state transition that could drastically change the accretion rate of the AGN. With cadenced surveys (e.g. starting in the 1990s with the {\it ROSAT} all-sky survey with a 6-month cadence, and to today, with optical time-domain surveys that scan the sky every few days), we are discovering more such extreme transients that indicate transformations in their accretion flow properties, perhaps due to the presence of stars in the nuclear regions, or even other black holes (both supermassive and stellar mass). 

\begin{figure*}
\begin{center}
\includegraphics[width=\textwidth]{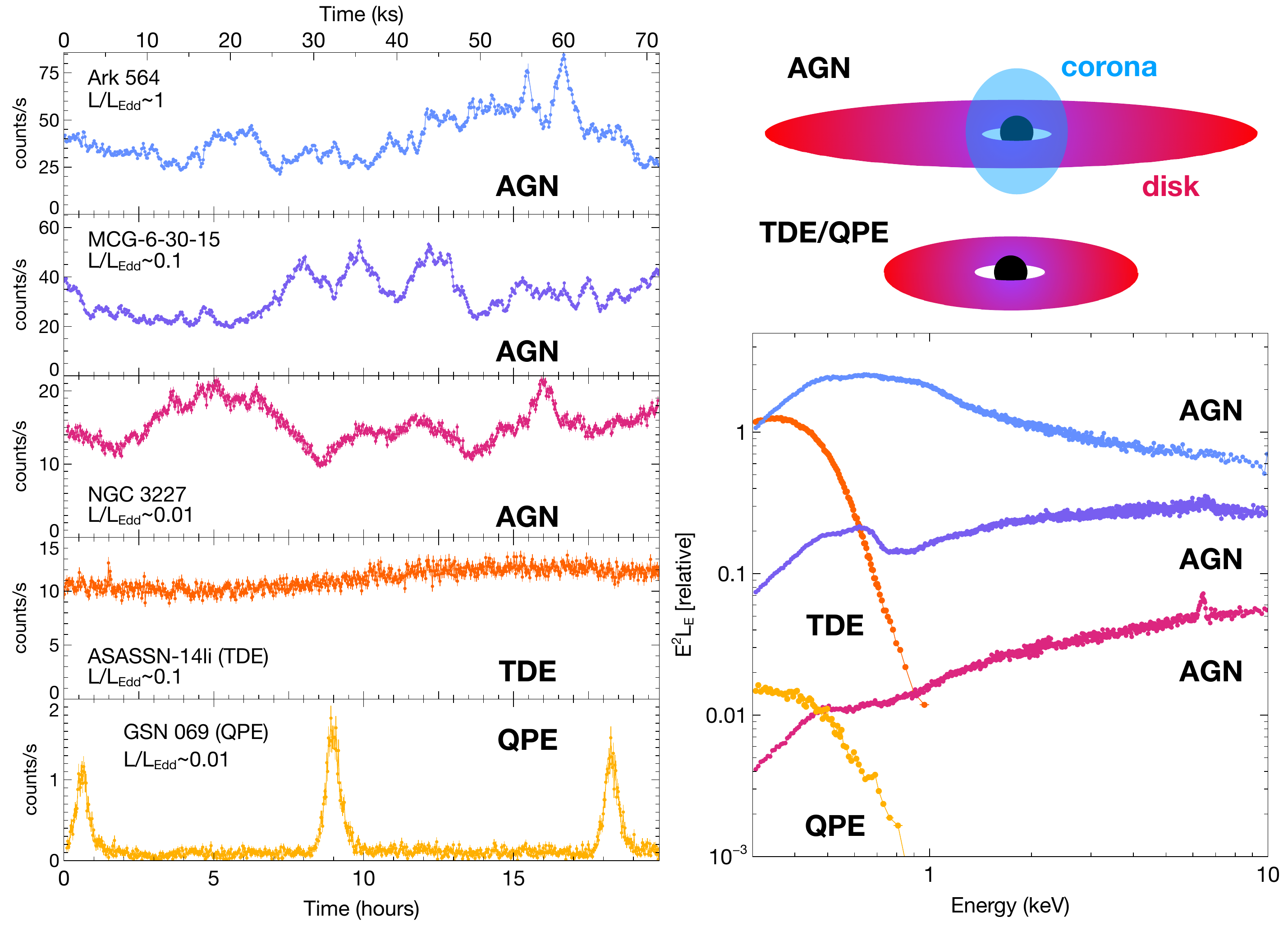}
\caption{A comparison of the X-ray variability and spectral properties of five $10^{6}$\msun~ accreting SMBHs, at different accretion rates. {\em Left:} The top three light curves demonstrate the stochastic nature of AGN light curves at $L/L_{Edd}\sim 1$, 0.1 and 0.01. The bottom two light curves are TDE candidates that do not follow the same variability patterns. In particular, the TDE candidate GSN~069 shows strong QPEs every $\sim 9$ hours. {\em Right:} The corresponding spectra for each of these source classes, scaled by their relevant luminosities. All standard AGN show hard X-ray components $>2$~keV. The TDE and QPE sources show no hard X-rays above $\sim 1$~keV, which is largely representative of so-called ``thermal" TDEs (Sec.~\ref{sec:tde}) and QPEs (Sec.~\ref{sec:QPE}). }
\label{fig:Mdot}
\end{center}
\end{figure*}

In this Section, we discuss some of the known and theorized nuclear transients, and what they can tell us about how black holes grow and affect their environments. We review what sets them apart from the standard AGN described in the previous section (see Fig.~\ref{fig:Mdot} for a comparison).
Finally, we describe what is possible with higher signal-to-noise X-ray observations.

\subsection{Tidal Disruption Events}\label{sec:tde}

A Tidal Disruption Event (TDE) occurs when a star approaches too closely to a SMBH to the point where the gravitational tidal forces exceed the self-gravity holding the star together. This process tears the star apart into a stream of stellar debris, some of which is gravitationally bound to the black hole, and forms an accretion disk around it. See \citet{Gezari2021} for a much more in-depth review of their multi-wavelength properties. Here, we focus on the X-ray properties of thermal TDEs (i.e. non-jetted), what they can teach us about accretion disk/corona physics, and on a new class of repeating TDEs that share some commonalities with X-ray Quasi-Periodic Eruptions (QPEs; Section~\ref{sec:QPE}).

\subsubsection{TDEs in X-rays}

The first TDE discoveries were made in the 1990s with the \textit{ROSAT} X-ray observatory \citep{bade1996,komossa2015}. More recently, its successor, \textit{eROSITA}, has discovered many more, with tens of confirmed X-ray selected TDEs (\citealt{sazonov21}), and several hundred candidates \citep{grotova25}. These events are identified as sources with no prior AGN activity (in X-ray or optical) that show sudden, significant X-ray brightening. The X-rays typically peak at $L_x > 10^{43} \, \mathrm{erg/s}$ on timescales of $<6$ months (the survey cadence) and decay over years, following roughly a $t^{-5/3}$ power-law. Despite no selection based on spectral shape, these transients consistently exhibit unusually soft, thermal X-ray spectra, matching the predictions of \citet{rees1988} that TDEs would emit soft X-rays from a $\sim 10^{6} \, \mathrm{K}$ accretion disk formed by the circularization of stellar debris. 

 In the past decade, most TDEs have been discovered in optical time-domain surveys, like ASAS-SN and the Zwicky Transient Facility. Their optical SEDs reveal photosphere temperatures of $\sim 10^{4} \, \mathrm{K}$, much cooler than a newly formed disk around a $\sim10^{6}$~\msun\ black hole \citep{yao2023}. The origin of the optical emission remains debated, with two leading interpretations: (1) X-rays from the disk are absorbed and reprocessed to longer wavelengths \citep{roth2016} by an envelope of unbound debris \citep{Metzger2016} or an optically thick disk wind \citep{dai18}, or (2) colliding debris streams produce shocks at large radii \citep{piran15}.
 
Initially, optically-selected TDEs were thought to rarely show X-ray emission, but extended X-ray campaigns have revealed that some TDEs only become X-ray bright hundreds of days after the optical peak \citep{gezari2017}. In fact, in a systematic study, \citet{Guolo2024xray} found that $\sim 40\%$ of optically selected TDEs show X-rays at some point, with most exhibiting delayed X-ray rises 100--200 days post-peak. This delay was initially interpreted as evidence for a late-forming accretion disk, supporting the debris-stream collision model for the optical emission \citep{gezari2017}. However, even if the disk forms promptly, X-rays may only emerge once the reprocessing layer becomes transparent as the accretion rate declines \citep{thomsen2022}. Regardless of debates on the emission mechanism in the first weeks to months, it is generally agreed that as the fallback rate decreases (over several months to years), the accretion rate on to the black hole will also decrease. This can manifest as a viscously spreading disk that persists much longer than expected from theoretical fallback rate extrapolations \citep{vanvelzen2019,mummery2020}.

\subsubsection{Where is the X-ray corona in TDEs?}

Notably, while hard X-rays ($>2$~keV) are a nearly ubiquitous feature of accreting AGN, TDEs, which achieve AGN-like luminosities, typically do not exhibit hard X-ray coronae \citep{komossa15}. This suggests that, {\em similar to black hole X-ray binaries, the accretion rate alone does not determine the production of a corona}. Possibly, a critical magnetic flux is required to build up as matter inflows through the disk to eventually trigger the formation of a corona \citep{begelman2014}. As the observed rates of TDEs have increased, and with it extensive X-ray follow-up observations with {\nicer}, {\swift}/XRT, {\xmm}, and {\chandra}, we have been able to monitor these events for years after their peak outbursts (e.g. \citealt{jonker2020}). These observations have led to the detection of the late-time formation of X-ray coronae in some TDEs \citep{kara2018,jonker2020,Guolo2024xray}. Moreover, in a few sources with well-monitored X-rays for several weeks to years post-peak, there have been observed ``state transitions" where the source changes from disk-dominated to corona-dominated and sometimes back again \citep{wevers2021,yao2022}. Some authors have speculated that these share some properties with the regular state transitions in black hole low-mass XRBs, but as of yet, there is no clear one-to-one mapping of TDE spectral evolution to the ``Q-diagrams" of low-mass XRBs.
\begin{marginnote}[]
    \entry{Q-diagram}{
    X-ray intensity vs. hardness ratio of outbursting XRBs, typically tracing a ``q"~shape that indicates hysteresis across accretion states.
    }
\end{marginnote}
Of course, it is important to remember that accounting for the differences in mass, a typical X-ray binary state transition lasting a few days is the equivalent of thousands of years for a supermassive black hole that is $10^{5}$ times more massive. Additional theoretical work is required to reconcile these disparate timescales if indeed they are manifestations of the same underlying physics.

Thanks to this common lack of a corona in TDEs, \citet{mummery2024} point out that TDEs may be the most ideal astrophysical systems for probing intrinsic small-scale disk turbulence because we have a clear view of the Wien tail from the innermost radii, whereas the Wien tail in typical AGN is either in the hidden EUV part of the SED (Section~\ref{sec:spectral}.1) or is ``contaminated" by the hard X-ray corona (Section~\ref{sec:spectral}.2). Therefore, future studies of detailed follow-up of the X-ray variability across a range of TDEs with different viewing angles and accretion rates (and disk scale heights) will probe the disk structure and dynamics, and perhaps also probe how and when magnetic flux can accumulate to produce a corona. {\em As such, while there are clear differences in how AGN and TDEs are fed, these new discoveries of TDEs as short-lived accretion episodes may help us understand more generally, how stable X-ray coronae in standard AGN are formed and powered.}

\subsubsection{New discovery space: IMBHs, Recoiling SMBHs, TDE in AGN \& Repeating TDEs}
\label{sec:pTDE}

To date, we know of roughly 150 confirmed TDEs \citep{Gezari2021, sazonov21,masterson24}. With this large and growing sample, we have begun to identify characteristics for determining bona fide TDEs beyond simply discovering nuclear flares with a power-law decline in previously quiescent galaxies. This allows us to extend our searches of TDEs, opening new discovery spaces. For instance, because we know that many TDEs have super-soft X-ray spectra, strong high-ionization lines like He II, or Bowen fluorescence lines (e.g. \citealt{vanvelzen2019}), we can begin to loosen requirements that bona fide TDEs exist in the nuclei of galaxies. This could allow for the discovery of TDEs from Intermediate Mass Black Holes (IMBHs; \citealt{maksym14}) or recoiling SMBHs \citep{ricarte21}. We can also loosen the requirement that the pre-existing galactic nucleus be quiescent. Given that processes like galaxy mergers can drive both stars and gas towards the central black hole, we should expect that some TDEs will occur in AGN. This has led to the discovery of several TDE-in-AGN candidates \citep{blanchard17,ricci20}, some of which have been labeled as ``Changing-Look AGN'' (see next Section~{\ref{sec:CLAGN}}).  

We can also begin to loosen requirements on the light curve behavior to search for repeated activity from a TDE. Such repeated TDEs are expected theoretically: only stars where the pericenter of its orbit is smaller than the tidal disruption radius will be destroyed, but many stars will have slightly larger pericenter distances, and in these systems the outer envelope of the star can get stripped off, causing a partial TDE (pTDE) \citep{guillochon13,coughlin19}. 
The rate of pTDEs could be orders of magnitude higher than that of complete TDEs \citep{stone16}, and have been theorized to contribute significantly to the growth of SMBHs \citep{bortolas23}. If the stellar remnant follows an elliptical orbit after a pTDE, it may be disrupted again after returning to pericenter, leading to repeating partial TDEs.

Observationally, the prototype object in this emerging source class of repeating pTDEs is ASASSN-14ko \citep{payne21} that shows repeated optical and X-ray flares every 114 days. Several other repeating events have recently been observed over timescales of months to years \citep{wevers23, malyali23, somalwar23,Evans2023}. It remains unclear how a partially disrupted star dissipates enough kinetic energy to be bound to the black hole with a period as short as $\sim100$~days. This has sparked a flurry of new theoretical ideas, including the break-up of stellar binaries \citep{cufari22,lu23} or the presence of SMBH binaries \citep{melchor24}. Whatever their origin, these new repeating and pTDEs offer a new way to probe the types of stars that are disrupted, and provide a broader understanding of the dynamics of stars in galactic nuclei.

\begin{figure*}
\begin{center}
\includegraphics[width=\textwidth]{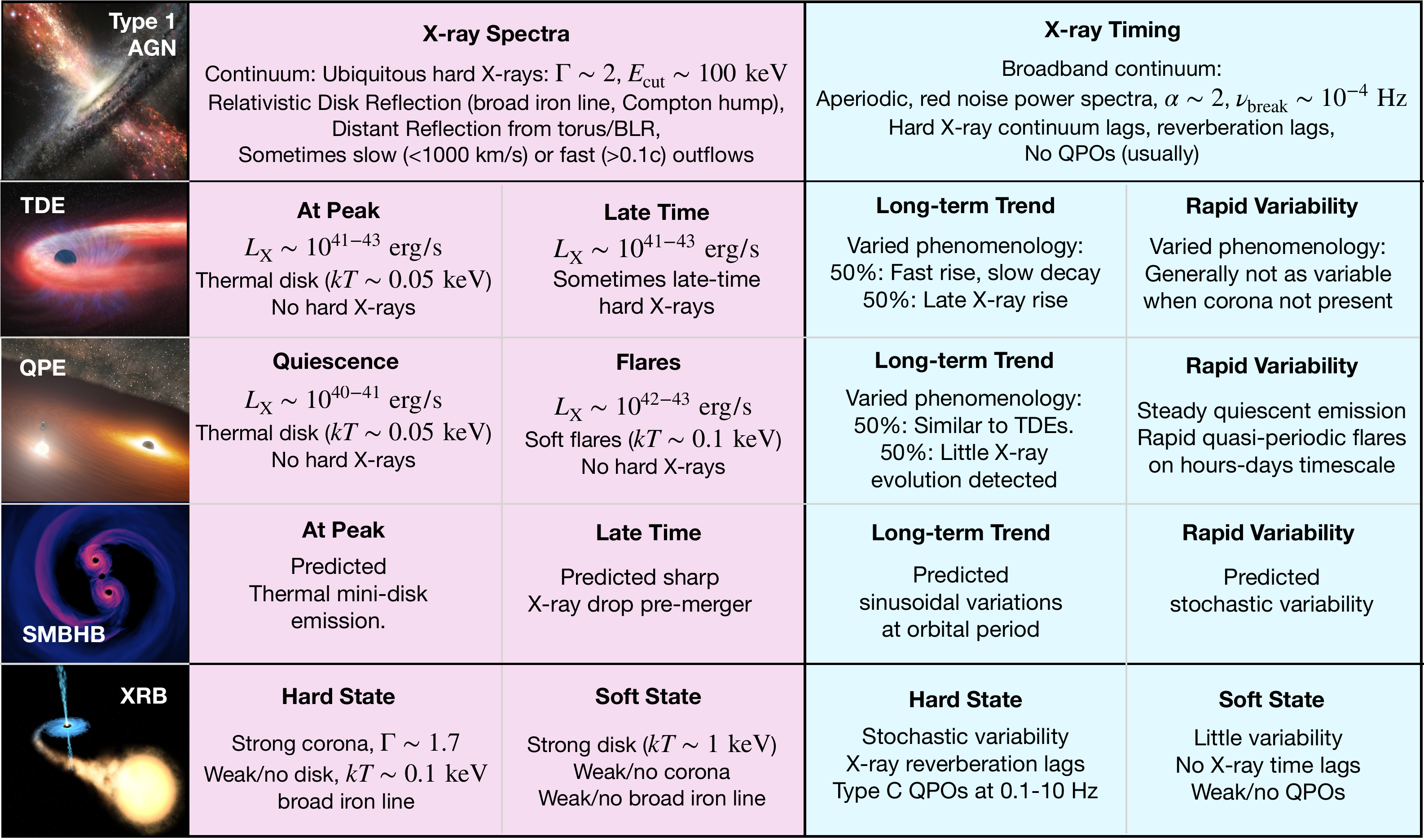}
\caption{Table summarizing some of the key spectral and timing properties of the different source types described in this review article. Accretion is by its very nature stochastic, and so there are exceptions to nearly all of these properties, but we attempt to summarize as a way to gain some traction on the vast phenomenology.}
\label{SummaryTable}
\end{center}
\end{figure*}

\subsection{Changing-Look AGN}\label{sec:CLAGN}

Concurrent with the discovery of TDEs was the discovery of other extreme variability phenomena occurring in the centers of active galaxies: so-called ``Changing-Look" (CLAGN)\footnote{The term CLAGN can refer both to AGN with variability due to accretion-rate changes (sometimes specified as ``Changing-State" AGN), and those where the variability is due to obscuration along the line-of-sight (so-called ``Changing-Obscuration" AGN). Here we use the term CLAGN to discuss only those that are definitively changing in accretion rate. See recent reviews by \citealt{Ricci+2023,komossa20} for more details and perspectives, and reviews of changing-obscuration events.} \citep{matt2003}, where AGN are seen to dramatically change in bolometric luminosity (beyond the stochastic variability discussed in Section~\ref{sec:timing}) on timescales of several months to years---faster than the expected viscous timescale at the outer accretion disk (See Eq.~\ref{eq:visc})\footnote{The viscous timescale at $\sim 10^3 R_g$ for a $10^8$\msun\ AGN is roughly $t_\mathrm{visc}\sim 500$\,years.}. 

Whereas the study of TDEs had the benefit of starting from a strong theoretical foundation (one predicted strong, fast flare rising to AGN-like luminosities in {\em previously quiescent, low-mass galaxies}), many other supermassive black holes exhibited extreme variability that {\em did} occur in systems with previous AGN activity, and therefore could not definitively be characterized as TDEs. 

The phenomenology of CLAGN is complex and likely indicates that their origin is also varied. They have been posited to be due to stars in the nuclear environments interacting with the AGN disk \citep{mckernan2022,Chen2024}, binary supermassive black holes \citep{Wang2020}, extreme {\it in situ} disk instabilities \citep{raj+nixon2021}, or due to reprocessing of a highly variable X-ray/EUV source \citep{lawrence2018,noda2018}. As such, they have garnered much attention, but despite knowing of these events for over two decades, this is a field much in its infancy because of the complex and heterogeneous phenomenology.  It is likely that in years to come, as larger all-sky and panchromatic observatories come online, this coarse description of extreme AGN variability will splinter into several new and exciting sub-fields. 

\subsubsection{Optically-selected CLAGN}

Early optical papers defined CLAGN as those that showed the (dis)appearance of the broad emission lines, indicating a transition between Type~1 and Type~2 AGN \citep{macleod2016}. The appearance of Type~1 broad lines was accompanied by a significant rise in the blue continuum (indicating a rise in the thermal emission from the accretion disk). The dramatic change in broad emission lines tells us that the unseen far-UV must be changing even more dramatically than the optical continuum, thus confirming their extreme nature \citep{lawrence2018}. Recent SDSS-V spectroscopic studies suggest that $\sim 0.4$\% of AGN may be CLAGN, with this fraction increasing by a few per cent in systems of lower mass \citep{Zeltyn2024}.

In recent years, there has been several efforts to identify CLAGN via optical photometric observations alone. Both photometric and spectroscopic approaches require care, as even stationary red noise processes can lead to one-off high-amplitude flares and do not necessarily indicate a physical change to the system (see e.g. \citealt{uttley2005} for a discussion of seemingly dramatic flares that arise from random realizations of a uniform physical process, in the context of black hole XRBs). These systematic statistical approaches (e.g. \citealt{Lopez-Navas2023, wang2024-clagn}) are important for developing homogeneous samples, and will become even more vital in the {\it Rubin} era, as our sample sizes increase.

\subsubsection{The X-ray properties of CLAGN}

The X-ray properties of CLAGN will be an important piece of the puzzle for understanding their nature, as the X-rays probe the accretion flow and magnetic field structure at the smallest scales. As of yet, there is no clear consensus on what happens to the X-rays in optically-selected CLAGN. In many cases, the X-rays track the observed optical variations (e.g. \citealt{krumpe2017}), and in other well-known cases the X-rays do not correlate with the optical. Perhaps the best known case of such phenomenon is 1ES~1927-654 \citep{trakhtenbrot2019}, where the optical increased by a factor of a few, followed by variations up and down in the X-rays by 4 orders of magnitude, including the disappearance and re-building of the hot X-ray corona. Such varied phenomenology in the X-rays is telling us something, and likely indicates that these extreme optical flares do not all have the same origin. 

Future optical discoveries should be accompanied by high-cadence X-ray monitoring (e.g. every few days) to characterize and track the often exotic evolution of the accretion flow structure on the smallest scales. Moreover, we need more real-time discoveries of CS-AGN in the X-rays to understand which comes first: is there a restructuring of the X-ray emitting inner accretion flow (where the viscous timescale is short, on the order of days) that---due to reprocessing---drives the optical variability that we observe, or is the CLAGN a much more global phenomenon, affecting the outer accretion disk (in the optical), which propagates inwards, thus later affecting the X-ray emitting region. These kinds of coordinated campaigns, multi-wavelength follow-up are vital for understanding the origin of CS-AGN.

\subsection{Quasi-Periodic Eruptions}\label{sec:QPE}

Quasi-Periodic Eruptions (QPEs) are an X-ray phenomenon occurring at the centers of galactic nuclei that have recently taken the astronomical community by storm. Discovered in 2019 (and announced rather epically by Giovanni Miniutti at an X-ray conference in Bologna to the shock of many audience-members, including these authors), there are now more theory papers trying to explain this perplexing phenomena than there are sources that show QPE behavior! QPEs are dramatic X-ray flashes occurring every few hours in a very regular (though not completely periodic) way. At the time of writing this review article, there are 9 confirmed QPEs (\citealt{miniutti2019, Giustini2020,arcodia2021,arcodia2024-ero34,nicholl24},  Hernandez-Garcia et al., in press, \citealt{chakraborty25}) and 2 candidates that show similar properties, but have only shown $\sim 2$ flares \citep{chakraborty2021,quintin2023,bykov2024}. 

\subsubsection{QPE observables}

The first QPE was discovered serendipitously, while examining the X-ray light curves of a peculiar nuclear transient, GSN~069 \citep{miniutti2019}. The other confirmed and candidate QPE sources show similar behavior. They have typical peak luminosities of $10^{42-43}$~erg/s, and typical recurrence times of several hours to days. They show soft X-ray spectra (with no hard X-ray corona!). In quiescence, they are well modeled by thermal accretion disks with $kT\sim 50-80$~eV, and during flares, they universally show hotter thermal-like spectra ($kT\sim 100-250$~eV), with the characteristic temperature hysteresis suggesting an expending emitting region during the flare \citep{miniutti2019}.  Some show very regular flaring (even a beautiful repeating long-short behavior of the flares; e.g. \citealt{Arcodia2024-ero2}), while others are less regular in their recurrence times \citep{Giustini2020,Chakraborty2024}. All occur in the nuclei of low-mass galaxies, in systems with masses ranging from $10^{5.5-6.5}$\msun, but as of yet, we do not see clear correlations in the recurrence time or flare duration with the black hole mass. 

More than half of the known QPE candidates take place in known TDE or CLAGN, and all (so far) occur several years after the peak luminosity (\citealt{miniutti2019,chakraborty2021,quintin2023, nicholl24}). Moreover, the host galaxies of QPEs and TDEs both show an over-representation in post-starburst galaxies \citep{wevers2022}. Interestingly. their optical spectra show no broad lines (suggesting there is intrinsically no BLR in these systems), though many do have narrow high-ionization lines, consistent with some past AGN-like activity \citep{wevers2022}. This is corroborated by recent MUSE studies that reveal that there is an over-representation of extended emission line regions in these systems, suggesting ionization from a previous, short-lived SMBH flare \citep{wevers2024}. 

It is now incontrovertible that some---if not all---QPEs occur in TDEs, but we cannot yet say that they {\em must} occur in TDEs. In standard AGN, where there are strong coronae that show multi-frequency red noise (Section~\ref{sec:timing}), it would be harder to convincingly pick out a $\sim 1000$~s quasi-periodic flare above the stochastic red noise background, and so physical models that predict QPEs in both TDEs and AGN are not currently ruled out by observations. 

\begin{figure*}
\begin{center}
\includegraphics[width=\textwidth]{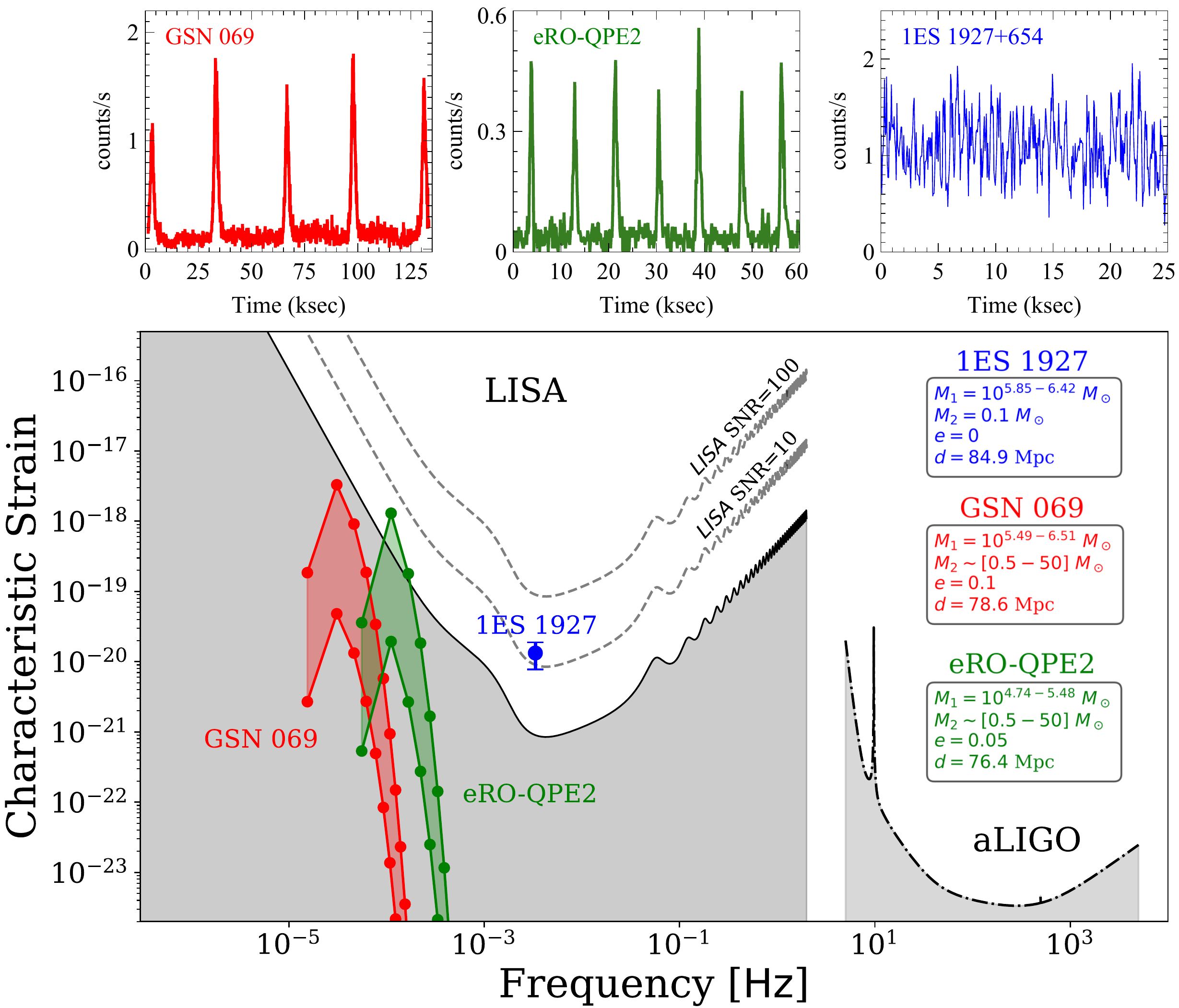}
\caption{{\em Top:} Example light curves of two Quasi Periodic Eruptions, GSN~069 (red) and eRO-QPE2 (green), which show periodic flaring on timescales of hours, and the X-ray light curve of 1ES~1927+654 (blue) that shows a $\sim 10$-minute period. A leading model for QPEs is that they are due to the interaction of an Extreme Mass Ratio Inspiral with the accretion disk of the supermassive black hole it is orbiting. If the orbiter model is correct, QPEs will be potential LISA gravitational wave sources. {\em Bottom:} The LISA sensitivity curve (shaded gray region) highlights the frequency range where LISA can detect signals, with dashed lines indicating signal-to-noise ratios (SNR) of 10 and 100. Overplotted are the predicted gravitational wave signals for GSN 069 (red), eRO-QPE2 (green), and 1ES 1927 (blue) based on the mass, eccentricity, and distance parameters, assuming an EMRI model. The ranges shown for GSN~069 and eRO-QPE2 reflect our uncertainty in the origin of the orbiter, whether it is a 0.5\msun\ star or a 50\msun\ black hole. The authors thank Joheen Chakraborty (MIT) for producing the LISA characteristic strain plot, and Megan Masterson (MIT) and Riccardo Arcodia (MIT) for the light curves.}
\label{fig:LISA}
\end{center}
\end{figure*}

\subsubsection{Models for the origin of QPEs}

There are two general flavors of models for the origin of QPEs: accretion disk instabilities or the interaction of an orbiting low-mass object (star or black hole) with the central SMBH. 

The disk instabilities models find inspiration in observations of some exotic black hole low-mass X-ray binaries like GRS~1915+105 and IGR~J17091-3624, which are known to show highly structured dramatic flaring light curves on a range of timescales from hundreds of seconds down to sub-second \citep{belloni2000,Altamirano2011,wang2024}. These have been explained by radiation pressure instabilities that cause a narrow ring in the accretion flow to oscillate between stable and unstable modes of accretion as the ring heats and cools \citep{Sniegowska2020}, or by warped accretion disks that can tear and break into discrete rings that accrete at timescales faster than the local viscous time \citep{raj+nixon2021}. 

More widely explored are the orbiter models, where QPEs have been proposed as due to repeated partial tidal disruption events \citep{king2022}, Roche-lobe overflow of a main-sequence star in a mildly eccentric orbit around the black hole \citep{lu2023,krolik2022,metzger2022} or stars/black holes (known as Extreme Mass Ratio Inspirals or EMRIs) colliding with a pre-existing disk, or a disk formed in a TDE \citep{linial2023,Franchini2023}. 
\begin{marginnote}[]
    \entry{EMRI, Extreme Mass Ratio Inspiral}{Typically refers to a stellar mass object orbiting a supermassive black hole.}
\end{marginnote}
The EMRI-disk collision models are particularly adept at explaining the observed long-short pattern observed in some QPEs, as the EMRI (star or compact object) on a mildly eccentric orbit passes twice per orbit through the accretion disk, each time shocking and ejecting material from the disk that expands throughout the flare. In this scenario, the cross section of the EMRI with the disk determines the amplitude of the flare. \citet{linial2023} find the cross section of a star is sufficient, while a compact object would need to be relatively high mass (e.g. a black hole with $\sim 50-100$~\msun) to have a Bondi radius large enough to explain the observed amplitudes. 

\subsubsection{How will we determine the origin of QPEs?} 

QPEs have the benefit of being regular, almost deterministic time series, that we can predict and model. The EMRI model---at its essence---allows for predictions for when and how long flares occur. Of course, these are not clean EMRI systems, as most numerical relativists have largely focused on, and the presence of an accretion disk (that may be warped or even precessing; \citealt{Franchini2023}) does significantly complicate the models, but in time, with long enough baselines, we should be able to put models to the test by predicting future flares. Already, multi-year monitoring programs are being used to test period modulations predicted by EMRI models \citep{Chakraborty2024, Arcodia2024-ero2, Miniutti24}.

The incontrovertible proof that QPEs are associated with EMRIs orbiting a SMBH would come from the joint detection of a QPE with a gravitational wave signal from LISA. \citet{Arcodia2024-ero2} estimate that if QPEs are caused by EMRIs, then those with $2-20$~hr flare recurrence times (of which there are currently 3~known) would emit GWs in the frequency range of $10^{-5}$ to $10^{-4}$~Hz, which is on the lower end of the frequency range where LISA is sensitive (Fig.~\ref{fig:LISA}). If the orbiter is a $40-100$\msun black hole, LISA could detect strains of  $10^{-18}$ to $10^{-17}$ at an SNR of $\sim 1$ for known QPEs. If the orbiter is a star, the known QPEs would be undetectable. 
Most recently, we discovered that another similar nuclear transient, 1ES~1927+654, exhibits even higher frequency quasi-periodic flaring that evolve over two years from an 18~min period down to 7~min \citep{masterson25}. If this source is also associated with an EMRI, it will emit gravitational waves at mHz frequencies, where LISA is most sensitive, and would be detected at an SNR~$\sim 10$.

While it remains to be seen whether the currently known QPEs will be LISA sources, they nonetheless inform our understanding of the dynamics of stars in galactic nuclei. Many models proposed for QPEs can also account for longer-period nuclear transients (e.g., \citealt{Evans2023,Guolo2024-period,Pasham2024}) and repeating partial TDEs in Section~\ref{sec:pTDE}), and these same models predict even closer encounters that would be detectable by LISA \citep{linial2023,metzger2022,melchor24}.
Given the growing number of QPE sources, repeating pTDEs and in light of recent discoveries of mHz oscillations around supermassive black hole transients \citep{pasham19,masterson25}, we consider it a hopeful prospect for future joint X-ray+LISA EMRI detections. This joint detection would allow for unprecedented understanding of accretion disks (as angular momentum losses from gas drag can dominate gravitational wave emission, and are extremely sensitive to the disk structure and surface density), as well as precision cosmology measurements.

\subsection{Supermassive black hole binary candidates}\label{sec:SMBHB}

One of the key predictions of hierarchical structure formation is that supermassive black holes (SMBHs) grow through successive mergers following the coalescence of their host halos. These SMBH mergers emit gravitational waves, detectable by Pulsar Timing Arrays (PTAs) for black holes in the range of $10^8$\msun\ to $10^{10}$\msun, and by LISA for those in the range of $10^5$\msun\ to $10^7$\msun. The recent NANOGrav PTA discovery of a potential gravitational wave background from a population of SMBHBs (\citealt{PTA}) and ESA's approval of the LISA mission for a 2035 launch (\citealt{LISA}) highlight the growing likelihood of such detections. While no definitive sub-parsec SMBHB has been identified to date, ongoing advancements in theoretical predictions of electromagnetic signatures continue to drive observational searches (\citealt{Bogdanovic2021}).

\subsubsection{Theoretical predictions of EM emission from SMBHBs}

The gas dynamics around SMBHBs and the subsequent electromagnetic signatures are highly uncertain, but it is likely that LISA and PTA-detectable, merging SMBHBs will have accretion disks, as they result in the interaction of gas-rich galaxies. It is generally agreed that SMBHB systems with large mass ratios exist in a circumbinary disk \citep{pringle91,macfadyen08}, which feeds mini-disks around the two black hole \citep{ryan17}. The circumbinary disk is expected to emit in optical/UV, while mini-discs will produce thermal emission in soft X-rays and a hard X-ray continuum at luminosities close to the Eddington limit \citep{dascoli18}. It then follows that as the black holes orbit each other, one should observe periodic luminosity fluctuations at the orbital frequency (i.e. at half the gravitational wave frequency). This signature should be strongest in the X-rays, that are dominated by mini-disk emission, and where strong Doppler beaming effects can enhance the variability amplitude \citep{dorazio15}. 

Recent theory works have studied the EM signatures within a few days of merger and soon after merger, and while there remains much debate about the exact dynamics, it is generally predicted that there will be dramatic changes to the broadband SED at the time of merger. For instance, as the black holes get close to merger, they will eventually lose contact with the circumbinary disk, thus depleting the gas reservoir to the mini-disks, which may result in a drop of the X-ray emission right before merger \citep{farris2015,dittmann2023}, perhaps by several orders of magnitude \citep{krauth2023}. Moreover, the anisotropic GW emission at merger will result in sudden mass-loss of the black hole (perhaps by a few per cent; \citealt{Tichy2008}) and a recoil kick of a few hundred km/s \citep{baker2008}, which can cause the circumbinary disk to fill in, and rebrightening to a single AGN (e.g. \citealt{Milosavljevic2005}).

\subsubsection{The ongoing quest for milli-parsec SMBHBs}

The advent of optical time domain surveys that have been running for $>10$ years have motivated several studies to systematically search for SMBHBs with orbital modulations of of a few years, corresponding to milli-pc separations (e.g., \citealt{graham2015,charisi2016}), though no unambiguous binaries have been found. Section~\ref{sec:timing} provides details for why this is such a tricky detection to make. Even single AGN are highly variable, on a range of timescales from hours to years. If one does not have a long enough baseline, or high enough cadence observations, it can be difficult to distinguish between ``normal" stochastic AGN variability and a periodic signal.  \cite{vaughan2016} examine this issue in detail, showing that AGN can easily have spurious ``phantom periods" (typically $\sim 3$ cycles). This happens more commonly in steep red noise power spectra (i.e. $\alpha > 2$), which happens to be what is observed in high-quality optical PSDs; \citealt{Mushotzky2011}). Moreover, it is known that the PSD shape varies with black hole mass and luminosity \citep{Arevalo2023}, and therefore it is vital to test against the correct null hypothesis that accounts for the diversity of PSD shapes seen. {\em In short, it is not sufficient to test for periods in optical time domain surveys against a null hypothesis of a Damped Random Walk.} 

\subsubsection{Future prospects for SMBHB discovery}

Fortunately, time-domain surveys are only getting better. Baselines are getting longer, and the quality of the light curves are improving. As such, with {\it Rubin/LSST} we have a better chance to robustly detect periodic light curves, but to truly decipher whether periodic behavior is due to merging black holes (rather than some other accretion disk instability or precession), we need to go beyond, and look for independent indicators that a binary is present. Of course, the joint detection of low-frequency gravitational waves would make this trivial, but until that day, there is still more we can do with electromagnetic observations. For instance, it would be hard to refute the existence of merging binary supermassive black holes if one also finds a negative period derivative consistent with gravitational wave radiation over many ($>10$) cycles. As the black holes inspiral, one might expect Doppler shifting of the X-ray emission from the mini-disks. If X-ray coronae form in mini-disks and irradiate the disks, one would expect double~peaked iron lines or ``rocking" iron lines moving from blue-to-red by $\sim 10,000$~km/s, at the orbital period. One could also search for reverberation lags from the mini-disks that are shorter than those expected from a single AGN with larger accretion disks (e.g. \citealt{liu2024}). Finally, as the black holes approach coalescence, we can look for X-ray dips, indicating the final detachment of the mini-disks from their circumbinary disks (e.g. \citealt{dittmann2023,krauth2023}). Regardless of the exact observables, the X-rays will be an important wavelength for searching for EM emission from SMBHBs, as these mergers are going to be gas-rich, and so the nucleus will likely be heavily extincted in the UV/optical, but bright in X-rays.

While the predicted {\it LISA} rates of binary supermassive black hole mergers is highly uncertain, we hope to observe several in the {\it LISA} era. Joint X-ray detections will provide vital information on the gas dynamics that enables merger, but they are also important for key gravitational wave studies as well: Gravitational waves are sensitive to the chirp mass, but as the EM observables are very sensitive to the mass ratio, a joint detection will nail down individual component masses. Moreover, gravitational waves alone have a degeneracy between distance and system inclination. X-ray fluorescence lines from the mini-disks can provide a robust measurement of the inclination (see Section~\ref{sec:reflection} for inclination constraints from reflection spectroscopy in single AGN). {\it A joint X-ray and GW detection of any binary SMBH candidate will allow us to break the distance-inclination degeneracy, thus enabling precision cosmology measurements beyond the local distance ladder.}
 
\section{SUMMARY \& OUTLOOK}\label{chap:summary}

In the last decade, a suite of X-ray missions, each with their own unique capabilities, has greatly expanded our understanding of the inner accretion flow closest to the event horizon. In short:
\begin{itemize}
\item As the first hard X-ray focusing mission, with {\nustar} we have an unprecedented view of the relativistically broadened iron~K and Compton hump features, which enables robust measurements of black hole spin in stellar mass and supermassive black holes (Section~\ref{sec:reflection}).
\item Long uninterrupted {\xmm} observations of AGN, and corresponding high-throughput {\nicer} observations of stellar mass black holes, have allowed us to measure the corresponding X-ray reverberation lags from light reflected off the accretion disk, solidifying the relativistic reflection interpretation and probing the geometry of the X-ray corona (Section~\ref{sec:reverberation}).
\item {\swift}, with its co-aligned X-ray and UV/optical optics, has enabled simultaneous reverberation mapping of the outer accretion disk, so we can probe the connection between the disk and corona, and how that correlates with energy released via accretion disk winds (Section~\ref{sec:outflows}).
\item Now with the recent launch of {\ixpe} in 2021 and {\xrism} in 2023, we have opened new windows into X-ray polarization (Section~\ref{sec:polarization}) and high-resolution X-ray spectroscopy (Section~\ref{sec:blr-torus}), respectively. It is likely that the biggest contributions of these missions are yet to come, as we collect more observations and models are developed to explain them. In fact, {\ixpe} is already opening new windows in our understanding of the geometry of the X-ray corona and {\xrism} in our understanding of large-scale gas flows in the BLR and the torus.
\end{itemize}

We are also living a renaissance in black hole accretion physics thanks to new time domain and multi-messenger facilities that are up-ending what we thought we knew about how supermassive black holes grow via accretion and mergers. We are witnessing previously quiescent black holes suddenly ``turn on" due to the tidal disruption of unfortunate stars on close-approach to the black hole (Section~\ref{sec:tde}); we regularly discover AGN that accrete material on timescales orders of magnitude faster than we thought theoretically possible (Section~\ref{sec:CLAGN}); we are seeing exotic quasi-periodic X-ray flaring in disks around supermassive black hole, that are posited to be due to the impact of a stellar mass object with the accretion disk (Section~\ref{sec:QPE}), and may lead to joint EM-GW discoveries with {\it LISA}. We hope to soon discover other {\it LISA} sources, too, with the robust discovery of supermassive black hole binaries at milli-parsec separations (Section~\ref{sec:SMBHB}), and regular joint detections of accreting SMBHs with high-energy neutrino events.

X-rays are a vital piece of this new time domain and multi-messenger view of supermassive black holes because they probe the regions closest to the event horizon. For instance,
\begin{itemize}
\item Although the origin of the optical emission in TDEs is not yet understood, the soft X-rays unambiguously originate from a small, newly formed accretion disk.
\item QPEs are an X-ray-only phenomenon, and to date, have no counterpart at longer wavelengths, which tells us that this phenomenon is also occurring close to the supermassive black hole, perhaps making some of them compelling {\it LISA} counterparts.
\item Binary supermassive black holes at milli-pc distances are thought to have mini-disks that emit in X-rays, with an optical/UV-emitting circumbinary disk. The strongest amplitude periodicity is predicted in the mini-disks. Furthermore, as these are likely gas-rich mergers, the nucleus could suffer from heavy extinction, but X-rays can penetrate.
\end{itemize}

Several other recent AGN discoveries not covered in this review would also be transformed by concurrent X-ray detections. For instance, {\it JWST} is now discovering distant high-$z$ quasar candidates ($z>6$) via rest-frame optical spectra of broad Balmer lines, but with little to no X-ray emission \citep{yue2024,maiolino2024}. Does the lack of X-ray emission mean they are mis-identified as quasars, or that the SED of these luminous systems is unlike typical Type~1 AGN in the local universe? A sensitive X-ray telescope that meets the horizon of {\it JWST} is needed. Such a concept, the Advanced X-ray Imaging Satellite (AXIS; \citealt{reynolds2023_AXIS}), is currently undergoing a NASA Phase A study for launch in 2032.  

Similarly tantalizing, {\it IceCUBE} is now detecting astrophysical high-energy neutrinos, some of which tentatively appear co-spatial with the two brightest Seyfert AGN in the X-ray band, NGC~1068 \citep{icecube} and NGC~4151 \citep{NGC4151_icecube}. Several studies suggest these neutrinos originate in the X-ray corona (e.g., \citealt{Inoue2019}). High-cadence X-ray monitoring of a large AGN sample, aimed at identifying correlations between coronal luminosity and neutrino flux, could provide definitive evidence for this link.

It is truly an exciting time to study supermassive black holes, and X-rays are an essential piece of the panchromatic view to understand their growth via accretion and mergers. With flagship missions {\em Athena} and {\em Lynx} slated to be launched in the late 2030s to 2050s, we require a generation of X-ray missions in the next decade to usher us in to this new era of supermassive black hole accretion and growth.  

\section*{DISCLOSURE STATEMENT}
The authors are not aware of any affiliations, memberships, funding, or financial holdings that might be perceived as affecting the objectivity of this review. 

\section*{ACKNOWLEDGMENTS}
This review has benefited abundantly from the contributions and inspiring discussions from many of our colleagues. We are particularly thankful to (in alphabetical order): 
Riccardo Arcodia,
Peter Boorman,
Joheen Chakraborty,
Saavik Ford,
Yan-Fei Jiang,
Tim Kallman,
Elias Kammoun,
Megan Masterson,
Guglielmo Mastroserio,
Barry McKernan,
Brian Metzer,
Richard Mushotzky,
Edward Nathan,
Scott Noble,
Christos Panagiotou,
Joanna Piotrowska,
Swati Ravi,
Dev Sadaula,
Navin Sridhar,
Phil Uttley, and
Marta Volonteri.

\bibliographystyle{ar-style2}
\bibliography{ref,jg_ref,my-references,2020_refs}

\section*{APPENDIX}

Extensive time-dependent photoionization calculations that include a great amount of microphysics (e.g.; elements other than hydrogen) have been presented by \cite{Rogantini+2022,Sadaula+2023,Luminari+2023} and \cite{Sadaula+2024}, with direct applications to astrophysical problems. Extending the discussion presented in 
Section~\ref{sec:timescales}, the relevant timescales for the interaction of radiation with matter in a photoionized gas are:

\begin{itemize}
    \item Photoionization time (PI),
    
    \begin{equation}
        t_\mathrm{PI} \sim \frac{1}{\gamma}
    \end{equation}
where $\gamma$ is the integrated radiation field $F(E)$ weighted by the photoabsorption cross section
$\sigma(E)_i$ of a given ion $i$:
\begin{equation}
    \gamma \simeq \int_{E_\mathrm{th}}^{1\mathrm{MeV}}{F(E)\sigma_i(E)dE/E}.
\end{equation}
From here it is clear that $t_\mathrm{PI}$ is mostly a function of the strength and overall shape of the local radiation field. The hydrogen photo-electric cross section can be approximated as $\sigma(E) = (E/E_\mathrm{th})^3\sigma(E_\mathrm{th})$, for energies above the ionization threshold $E_\mathrm{th}=13.6$\,eV, or zero otherwise; and $\sigma(E_\mathrm{th})=6.923\times10^{-18}$\,cm$^2$. In the case of disks illuminated by a X-ray corona, the radiation field can be approximated as power-law in energy $F(E)=F_0E^{-\alpha}$, and for typical Seyfert AGN $\alpha=1$ (note that $\Gamma=\alpha+1$, is the photon index, which is conventionally used by the X-ray community). One can then write a simple analytic approximation in the form:

\begin{equation}
    t_\mathrm{PI} \sim \frac{4 E_\mathrm{th}\ln{(1\mathrm{MeV}/E_\mathrm{th})}}{\sigma(E_\mathrm{th})F_0} \simeq 1.4\times10^8/F_\mathrm{x},
\end{equation}
where $F_\mathrm{x}$ is the net illuminating energy flux, $F_\mathrm{x} = \int{F(E)dE}$, which can be estimated as $F_\mathrm{x} = L_\mathrm{x}/4\pi R^2$. For typical AGN accreting at 10\% Eddington ($\lambda_\mathrm{Edd}$), and assuming a bolometric correction of $\kappa_\mathrm{x} = 0.1$ (see Section~\ref{sec:disk-corona}), the illuminating flux {\it at the disk surface} ranges  $F_\mathrm{x}\sim10^{15} - 10^{23}$\,erg\,cm$^{-2}$\,s$^{-1}$, and thus the PI times are always very short; $t_\mathrm{PI}\sim10^{-15} - 10^{-7}$\,s. These estimates will change for a plasma containing heavier atoms that can have a larger PI cross section, but are roughly consistent with the estimates reported by \cite{Sadaula+2023} (see their Fig.~1).

    \item Recombination time,

    \begin{equation}
        t_\mathrm{rec} \sim \frac{1}{n\alpha_r(T)}
        \label{eq:rec}
    \end{equation}
where $n$ is the gas density, and $\alpha_r(T)$ is the recombination rate, which is a rather complicated function of the temperature. Recombination rates are specific to a particular state in each ion, but generally they decrease with increasing temperature, making the recombination time larger for a given density. For the hydrogen recombination rate one can adopt the fitting formula from \cite{Badnell+2006},
\begin{equation}
    \alpha_r(T) = 8.32\times 10^{-11} \left[ \sqrt{T/2.97} (1+\sqrt{T/2.97})^{1-0.75} \times (1-\sqrt{T/7\times10^5})^{1+0.75} \right]^{-1}.
\end{equation}
This equation captures the complexity of the recombination rate in the $\sim 10^4 - 10^6$\,K temperature range. At $T\gtrsim10^6$\,K it can be approximated as $\alpha_r(T)\approx 10^{-4}T^{-1.7}$. A more general expression for $t_\mathrm{rec}$ that includes ions other than hydrogen is given by \cite{Nicastro+1999}.
\end{itemize}

\end{document}